\documentclass[12pt]{article}
\usepackage[cp1250]{inputenc}
\usepackage{amsmath,amssymb}

\def\MB{\mathbf}
\def\MC{\mathcal}
\def\MR{\mathrm}
\def\MF{\mathfrak}

\def\wh{\widehat}
\def\ol{\overline}

\def\Tr{\operatorname{\MR{Tr}}\nolimits}

\def\tsum{\textstyle\sum\limits}
\def\tprod{\textstyle\prod\limits}

\def\slacs#1{\setlength{\arraycolsep}{#1}}

\slacs{.4ex}
\begin{document}

\begin{center}
\Large\textbf{Nested Bethe Ansatz for RTT--Algebra of
$U_q\bigl(\MR{sp}(2n)\bigr)$ Type}
\end{center}

\centerline{\v{C}. Burd\'{\i}k\,$^{1)}$, O. Navr\'atil\,$^{2)}$}

\vskip5mm

\centerline{$^{1)}$\,Faculty of Nuclear Sciences and Physical Engineering, CTU,}
\centerline{Trojanova 13, Prague, Czech Republic}
\centerline{email: burdices@kmlinux.fjfi.cvut.cz}

\vskip3mm

\centerline{$^{2)}$\,Faculty of Transportation Sciences, CTU,}
\centerline{Na Florenci 25, Prague, Czech Republic}
\centerline{email: navraond@fd.cvut.cz}

\begin{abstract}
{We study the highest weight representations of the
RTT--algebras for the R--matrix of $\MR{sp}_q(2n)$ type by the
nested algebraic Bethe ansatz. It is a generalization of our study
for R--matrix of $\MR{sp}(2n)$ and $\MR{so}(2n)$ type.}
\end{abstract}

\section{Introduction}
 \label{sec:1}

The formulation of the quantum inverse scattering method, or
algebraic Bethe ansatz, by the Leningrad school \cite{FST79}
provides eigenvectors and eigenvalues of the transfer matrix. The
latter is the  generating function of the conserved quantities of a
large family of quantum integrable models. The transfer matrix
eigenvectors are constructed from the representation theory of the
RTT--algebras. In order to construct these eigenvectors, one should
first  prepare Bethe vectors, depending on a set of complex
variables. The first formulation of the Bethe vectors for the
$\MR{gl}(n)$--invariant models was given by P.P. Kulish and N.Yu.
Reshetikhin in \cite{KR83} where the nested algebraic Bethe ansatz
was introduced. These vectors are given by recursion on the rank of
the algebra. Our calculation is some $q$--generalization of the
construction which we published in recent works \cite{BNSP},
\cite{BNSO} and \cite{BuNa} for the non-deformed case of
$\MR{sp}(2n)$, $\MR{so}(2n)$ and $\MR{sp}(4)$. Our construction of
Bethe vectors used the new RTT--algebra $\tilde{\MC{A}}_n$ which is
defined in section 3 and is not the RTT--subalgebra of
$\MR{sp}_q(2n)$ type. This algebra has two RTT--subalgebras of
$\MR{gl}_q(n)$ type and the study of the nested Bethe ansatz for
this RTT--algebra is in progress. The simplest case for $n=2$ was
really solved and we will publish it in the next paper. Our
construction of Bethe vectors is in any sense a generalization of
Re\-she\-ti\-khin's results \cite{R85}. Another approach to the
nested Bethe ansatz for very special representations of
RTT--algebras of $\MR{sp}(2n)$ type was given by Martin and Ramas
\cite{MaRa}.

The proofs of many claims are explained from the reason of transparency
of the main text in the Appendix.

\section{Basic definitions and notation}
 \label{sec:2}

Let indices go trough the set $\{\pm1,\pm2,\ldots,\pm n\}$. We will
denote by $\MB{E}^k_i$ the matrices that have all elements equal to
zero with the exception of the element on the $i$-th row and $k$-th
column that is equal to one. Then relation
$\MB{E}^k_i\MB{E}^s_r=\delta^k_r\MB{E}^s_i$ is valid and
$\MB{I}={\tsum_{k=-n}^n}\MB{E}^k_k$ is the unit matrix.

We will consider the R--matrix of $U_q\bigl(\MR{sp}(2n)\bigr)$ which
has the shape
$$
\begin{array}{l}
\MB{R}(x)=
\dfrac1{\alpha(x)}
\Bigl({\tsum_{i,k;\,i\neq\pm k}}\MB{E}^i_i\otimes\MB{E}^k_k+
f(x){\tsum_i}\MB{E}^i_i\otimes\MB{E}^i_i+
f(x^{-1}q^{-n-1}){\tsum_i}\MB{E}^i_i\otimes\MB{E}^{-i}_{-i}+\\[9pt]
\hskip30mm+
g(x){\tsum_{k<i}}\MB{E}^i_k\otimes\MB{E}^k_i-
g(x^{-1}){\tsum_{i<k}}\MB{E}^i_k\otimes\MB{E}^k_i-\\[9pt]
\hskip30mm-
g(xq^{n+1}){\tsum_{k<i}}q^{k-i}\epsilon_i\epsilon_k\MB{E}^i_k\otimes\MB{E}^{-i}_{-k}+
g(x^{-1}q^{-n-1}){\tsum_{i<k}}q^{k-i}\epsilon_i\epsilon_k\MB{E}^i_k\otimes\MB{E}^{-i}_{-k}\Bigr)
\end{array}
$$
where $\epsilon_i=\MR{sign}(i)$ and
$$
f(x)=\frac{xq-x^{-1}q^{-1}}{x-x^{-1}}\,,\qquad
g(x)=\frac{x(q-q^{-1})}{x-x^{-1}}\,,\qquad
\alpha(x)=1+\frac{q-q^{-1}}{x-x^{-1}}\,.
$$
This R--matrix satisfies the Yang--Baxter equation
$$
\MB{R}_{1,2}(x)\MB{R}_{1,3}(xy)\MB{R}_{2,3}(y)=
\MB{R}_{2,3}(y)\MB{R}_{1,3}(xy)\MB{R}_{1,2}(x)
$$
and is invertible. Concretely, the inverse matrix has the form
$$
\begin{array}{l}
\MB{R}^{-1}(x)=
\dfrac1{\alpha(x^{-1})}
\Bigl({\tsum_{i,k;\,i\neq\pm k}}\MB{E}^i_i\otimes\MB{E}^k_k+
f(x^{-1}){\tsum_i}\MB{E}^i_i\otimes\MB{E}^i_i+
f(xq^{-n-1}){\tsum_i}\MB{E}^i_i\otimes\MB{E}^{-i}_{-i}-\\[9pt]
\hskip30mm-
g(x){\tsum_{k<i}}\MB{E}^i_k\otimes\MB{E}^k_i+
g(x^{-1}){\tsum_{i<k}}\MB{E}^i_k\otimes\MB{E}^k_i+\\[9pt]
\hskip30mm+
g(xq^{-n-1}){\tsum_{k<i}}q^{i-k}\epsilon_i\epsilon_k\MB{E}^i_k\otimes\MB{E}^{-i}_{-k}-
g(x^{-1}q^{n+1}){\tsum_{i<k}}q^{i-k}\epsilon_i\epsilon_k\MB{E}^i_k\otimes\MB{E}^{-i}_{-k}\Bigr)
\end{array}
$$

The RTT--algebra of $U_q\bigl(\MR{sp}(2n)\bigr)$ type  is an
associative algebra $\MC{A}$ with unit that is generated by
$T^i_k(x)$ for which the monodromy operator
$$
\MB{T}(x)={\tsum_{i,k=-n}^n}\MB{E}^k_i\otimes T^i_k(x)
$$
fulfills the RTT--equation
$$
\MB{R}_{1,2}(xy^{-1})\MB{T}_1(x)\MB{T}_2(y)=
\MB{T}_2(y)\MB{T}_1(x)\MB{R}_{1,2}(xy^{-1})\,.
$$
From the invertibility of the R--matrix we have that the operator
$$
H(x)=\Tr\bigl(\MB{T}(x)\bigr)= {\tsum_{i=-n}^n}T^i_i(x)
$$
fulfills the equation   $H(x)H(y)=H(y)H(x)$ for any $x$ and $y$.

\bigskip

We suppose that in the representation space $\MC{W}$ of the  RTT--algebra $\MC{A}$
there exists a vacuum vector $\omega\in\MC{W}$, for which
$\MC{W}=\MC{A}\omega$ and
$$
T^i_k(x)\omega=0\quad\MR{for}\quad i<k\,,\qquad
T^i_i(x)\omega=\lambda_i(x)\omega\quad\MR{for}\quad
i=\pm1,\,\pm2,\,\ldots,\,\pm n\,.
$$
In the vector space $\MC{W}=\MC{A}\omega$, we will look for
eigenvectors of $H(x)$.

\section{RTT--algebra $\tilde{\MC{A}}_n$}
 \label{sec:3}

In the RTT--algebra $\MC{A}$, we have the RTT--subalgebras
$\MC{A}^{(+)}$ and $\MC{A}^{(-)}$ that are generated by the elements
$T^i_k(x)$ and $T^{-i}_{-k}(x)$, where $i,\,k=1,\,2,\,\ldots,\,n$.
First, we will study the subspace
$$
\MC{W}_0=\MC{A}^{(+)}\MB{A}^{(-)}\omega\subset\MC{W}=\MC{A}\omega\,.
$$

\medskip

 \noindent
\underbar{\textbf{Lemma 1.}} For any  $i,\,k=1,\,2,\,\ldots,\,n$ and
any $\Omega\in\MC{W}_0$  $T^{-i}_k(x)\Omega=0$ is valid.

\bigskip

\underbar{\textbf{Lemma 2.}}
If we denote
$$
\MB{T}^{(+)}(x)={\tsum_{i,k=1}^n}\MB{E}^k_i\otimes T^i_k(x)\,,
\hskip10mm
\MB{T}^{(-)}(x)={\tsum_{i,k=1}^n}\MB{E}^{-k}_{-i}\otimes
T^{-i}_{-k}(x)\,,
$$
then relation
\begin{equation}
 \label{RTT-epsilon}
\MB{R}^{(\epsilon_1,\epsilon_2)}_{1,2}(xy^{-1})
\MB{T}^{(\epsilon_1)}_1(x)\MB{T}^{(\epsilon_2)}_2(y)=
\MB{T}^{(\epsilon_2)}_2(y)\MB{T}^{(\epsilon_1)}_1(x)
\MB{R}^{(\epsilon_1,\epsilon_2)}_{1,2}(xy^{-1})\,,
\end{equation}
is for any $\epsilon_1,\,\epsilon_2=\pm$ valid on the space $\MC{W}_0$,
where
$$
\begin{array}{l}
\MB{R}^{(+,+)}_{1,2}(x)= \dfrac1{f(x)}
\Bigl({\tsum_{i,k=1;\,i\neq k}^n}\MB{E}^i_i\otimes\MB{E}^k_k+
f(x){\tsum_{i=1}^n}\MB{E}^i_i\otimes\MB{E}^i_i+\\[9pt]
\hskip20mm+
g(x){\tsum_{1\leq k<i\leq n}}\MB{E}^i_k\otimes\MB{E}^k_i-
g(x^{-1}){\tsum_{1\leq i<k\leq n}}\MB{E}^i_k\otimes\MB{E}^k_i\Bigr)\\[12pt]
\MB{R}^{(-,-)}_{1,2}(x)= \dfrac1{f(x)}
\Bigl({\tsum_{i,k=1;\,i\neq k}^n}\MB{E}^{-i}_{-i}\otimes\MB{E}^{-k}_{-k}+
f(x){\tsum_{i=1}^n}\MB{E}^{-i}_{-i}\otimes\MB{E}^{-i}_{-i}+\\[9pt]
\hskip20mm+
g(x){\tsum_{1\leq i<k\leq n}}\MB{E}^{-i}_{-k}\otimes\MB{E}^{-k}_{-i}-
g(x^{-1}){\tsum_{1\leq k<i\leq n}}\MB{E}^{-i}_{-k}\otimes\MB{E}^{-k}_{-i}\Bigr)\\[12pt]
\MB{R}^{(+,-)}_{1,2}(x)=
{\tsum_{i,k=1;\,i\neq k}^n}\MB{E}^i_i\otimes\MB{E}^{-k}_{-k}+
f(x^{-1}q){\tsum_{i=1}^n}\MB{E}^i_i\otimes\MB{E}^{-i}_{-i}-\\[9pt]
\hskip20mm-
g(xq^{-1}){\tsum_{1\leq k<i\leq n}}q^{k-i}\MB{E}^i_k\otimes\MB{E}^{-i}_{-k}+
g(x^{-1}q){\tsum_{1\leq i<k\leq n}}q^{k-i}\MB{E}^i_k\otimes\MB{E}^{-i}_{-k}\\[12pt]
\MB{R}^{(-,+)}_{1,2}(x)=
{\tsum_{i,k=1;\,i\neq k}^n}\MB{E}^{-i}_{-i}\otimes\MB{E}^k_k+
f(x^{-1}q^{-n-1}){\tsum_{i=1}^n}\MB{E}^{-i}_{-i}\otimes\MB{E}^i_i-\\[9pt]
\hskip20mm-
g(xq^{n+1}){\tsum_{1\leq i<k\leq n}}q^{i-k}\MB{E}^{-i}_{-k}\otimes\MB{E}^i_k+
g(x^{-1}q^{-n-1}){\tsum_{1\leq k<i\leq n}}q^{i-k}\MB{E}^{-i}_{-k}\otimes\MB{E}^i_k\,.
\end{array}
$$

\bigskip

\underbar{\textbf{Proposition  1.}}
If we define
\begin{eqnarray*}
\tilde{\MB{R}}_{1,2}(x)&=&\MB{R}^{(+,+)}_{1,2}(x)+\MB{R}^{(+,-)}_{1,2}(x)+
\MB{R}^{(-,+)}_{1,2}(x)+\MB{R}^{(-,-)}_{1,2}(x)\\
\tilde{\MB{T}}(x)&=&\MB{T}^{(+)}(x)+\MB{T}^{(-)}(x)\,,
\end{eqnarray*}
we have on the space $\MC{W}_0$ the RTT--equation
$$
\tilde{\MB{R}}_{1,2}(xy^{-1})\tilde{\MB{T}}_1(x)\tilde{\MB{T}}_2(y)=
\tilde{\MB{T}}_2(y)\tilde{\MB{T}}_1(x)\tilde{\MB{R}}_{1,2}(xy^{-1})\,.
$$
Moreover, the R--matrix $\tilde{\MB{R}}(x)$ fulfills the
Yang--Baxter equation
$$
\tilde{\MB{R}}_{1,2}(x)\tilde{\MB{R}}_{1,3}(xy)\tilde{\MB{R}}_{2,3}(y)=
\tilde{\MB{R}}_{2,3}(y)\tilde{\MB{R}}_{1,3}(xy)\tilde{\MB{R}}_{1,2}(x)
$$
and has the inverse matrix
$$
\bigl(\tilde{\MB{R}}_{1,2}(x)\bigr)^{-1}=
\bigl(\MB{R}^{(+,+)}_{1,2}(x)\bigr)^{-1}+
\bigl(\MB{R}^{(+,-)}_{1,2}(x)\bigr)^{-1}+
\bigl(\MB{R}^{(-,+)}_{1,2}(x)\bigr)^{-1}+
\bigl(\MB{R}^{(-,-)}_{1,2}(x)\bigr)^{-1}
$$
where
$$
\begin{array}{l}
\bigl(\MB{R}^{(+,+)}_{1,2}(x)\bigr)^{-1}=\dfrac1{f(x^{-1})}
\Bigl({\tsum_{i,k=1;\,i\neq k}^n}\MB{E}^i_i\otimes\MB{E}^k_k+
f(x^{-1}){\tsum_{i=1}^n}\MB{E}^i_i\otimes\MB{E}^i_i-\\[9pt]
\hskip30mm-
g(x){\tsum_{1\leq k<i\leq n}}\MB{E}^i_k\otimes\MB{E}^k_i+
g(x^{-1}){\tsum_{1\leq i<k\leq n}}\MB{E}^i_k\otimes\MB{E}^k_i\Bigr)\\[12pt]
\bigl(\MB{R}^{(-,-)}_{1,2}(x)\bigr)^{-1}=\dfrac1{f(x^{-1})}
\Bigl({\tsum_{i,k=1;\,i\neq k}^n}\MB{E}^{-i}_{-i}\otimes\MB{E}^{-k}_{-k}+
f(x^{-1}){\tsum_{i=1}^n}\MB{E}^{-i}_{-i}\otimes\MB{E}^{-i}_{-i}-\\[9pt]
\hskip30mm-
g(x){\tsum_{1\leq i<k\leq n}}\MB{E}^{-i}_{-k}\otimes\MB{E}^{-k}_{-i}+
g(x^{-1}){\tsum_{1\leq k<i\leq n}}\MB{E}^{-i}_{-k}\otimes\MB{E}^{-k}_{-i}\Bigr)\\[12pt]
\bigl(\MB{R}^{(+,-)}_{1,2}(x)\bigr)^{-1}=
{\tsum_{i,k=1;\,i\neq k}^n}\MB{E}^i_i\otimes\MB{E}^{-k}_{-k}+
f(xq^{-n-1}){\tsum_{i=1}^n}\MB{E}^i_i\otimes\MB{E}^{-i}_{-i}+\\[9pt]
\hskip30mm+
g(xq^{-n-1}){\tsum_{1\leq k<i\leq n}}q^{i-k}\MB{E}^i_k\otimes\MB{E}^{-i}_{-k}-
g(x^{-1}q^{n+1}) {\tsum_{1\leq i<k\leq n}}q^{i-k}\MB{E}^i_k\otimes\MB{E}^{-i}_{-k}\\[12pt]
\bigl(\MB{R}^{(-,+)}_{1,2}(x)\bigr)^{-1}=
{\tsum_{i,k=1;\,i\neq k}^n}\MB{E}^{-i}_{-i}\otimes\MB{E}^k_k+
f(xq){\tsum_{i=1}^n}\MB{E}^{-i}_{-i}\otimes\MB{E}^i_i+\\[9pt]
\hskip30mm+
g(xq){\tsum_{1\leq i<k\leq n}}q^{k-i}\MB{E}^{-i}_{-k}\otimes\MB{E}^i_k-
g(x^{-1}q^{-1}){\tsum_{1\leq k<i\leq n}}q^{k-i}\MB{E}^{-i}_{-k}\otimes\MB{E}^i_k
\end{array}
$$

\underbar{\textsc{Proof.}} The validity of the RTT--equation is
Lemma 2. The Yang--Baxter equation, which is equivalent to the
equations
\begin{equation}
 \label{YB-epsilon}
\MB{R}^{(\epsilon_1,\epsilon_2)}_{1,2}(x)
\MB{R}^{(\epsilon_1,\epsilon_3)}_{1,3}(xy)
\MB{R}^{(\epsilon_2,\epsilon_3)}_{2,3}(y)=
\MB{R}^{(\epsilon_2,\epsilon_3)}_{2,3}(y)
\MB{R}^{(\epsilon_1,\epsilon_3)}_{1,3}(xy)
\MB{R}^{(\epsilon_1,\epsilon_2)}_{1,2}(x)
\end{equation}
and the conditions for the inverse $R$--matrix, i.e. the relations
$$
\MB{R}^{(\epsilon_1,\epsilon_2)}_{1,2}(x)
\bigl(\MB{R}^{(\epsilon_1,\epsilon_2)}_{1,2}(x)\bigr)^{-1}=
\MB{I}_{\epsilon_1}\otimes\MB{I}_{\epsilon_2}\,,\qquad
\MR{where}\quad \MB{I}_+={\tsum_{i=1}^n}\MB{E}^i_i\,,\quad
\MB{I}_-={\tsum_{i=1}^n}\MB{E}^{-i}_{-i}\,,
$$
can be shown by direct calculation.

\bigskip

\underbar{\textbf{Definition.}} We denote the RTT--algebra defined
by the R--matrix $\tilde{\MB{R}}(x)$ as $\tilde{\MC{A}}_n$.

\bigskip

We find out by the standard procedure from the RTT--equation
(\ref{RTT-epsilon}) that in the RTT--algebra $\tilde{\MC{A}}_n$
mutually commutate not only the operators $\tilde{H}(x)$ and
$\tilde{H}(y)$, where
$$
\tilde{H}(x)=\Tr_{(+,-)}\bigl(\tilde{\MB{T}}(x)\bigr)=
\Tr_+\bigl(\MB{T}^{(+)}(x)\bigr)+\Tr_-\bigl(\MB{T}^{(-)}(x)\bigr)=
{\tsum_{i=1}^n}\bigl(T^i_i(x)+T^{-i}_{-i}(x)\bigr)
$$
but also all operators $\tilde{H}^{(\pm)}(x)$ and
$\tilde{H}^{(\pm)}(y)$, where
$$
\tilde{H}^{(+)}(x)=\Tr_+\bigl(\MB{T}^{(+)}(x)\bigr)={\tsum_{i=1}^n}T^i_i(x)\,,
\qquad
\tilde{H}^{(-)}(x)=\Tr_-\bigl(\MB{T}^{(-)}(x)\bigr)={\tsum_{i=1}^n}T^{-i}_{-i}(x)\,.
$$

\section{General shape of eigenvectors}
 \label{sec:4}

Let $\vec{u}=(u_1,u_2,\ldots,u_M)$ be an ordered set of mutually
different complex numbers.
We will look for eigenvectors in the form
$$
\MF{V}(\vec{u})={\tsum_{i_1,\ldots,i_M,k_1,\ldots,k_M=1}^n}
T^{i_1}_{-k_1}(u_1)T^{i_2}_{-k_2}(u_2)\ldots T^{i_M}_{-k_M}(u_M)
\Phi^{k_1,k_2,\ldots,k_M}_{i_1,i_2,\ldots_{i_M}}
$$
where $\Phi^{k_1,k_2,\ldots,k_M}_{i_1,i_2,\ldots_{i_M}}\in\MC{W}_0$.
Let us denote
$$
\MB{B}(u)={\tsum_{i,k=1}^n}\MB{e}_i\otimes\MB{f}^{-k}\otimes
T^i_{-k}(u)\in \MC{V}_{+}\otimes\MC{V}^*_{-}\otimes\MC{A}
$$
where $\MB{e}_i$ is the basis of the space $\MC{V}_{+}$ and
$\MB{f}^{-k}$ is the basis of the space $\MC{V}^*_{-}$
and define
$$
\begin{array}{l}
\MB{B}_{1,\ldots,M}(\vec{u})=
\MB{B}_1(u_1)\otimes\MB{B}_2(u_2)\otimes\ldots\otimes\MB{B}_M(u_M)=\\[6pt]
\hskip10mm= {\tsum_{i_1,\ldots,i_M,k_1,\ldots,k_M}^n}
\MB{e}_{i_1}\otimes\ldots\otimes\MB{e}_{i_M}\otimes
\MB{f}^{-k_1}\otimes\ldots\otimes\MB{f}^{-k_M}\otimes
T^{i_1}_{-k_1}(u_1)\ldots T^{i_M}_{-k_M}(u_M)
\end{array}
$$
If $\MB{f}^r$ is the dual basis with respect to $\MB{e}_i$ in the
space $\MC{V}^*_{+}$ and $\MB{e}_{-s}$ is the dual basis with
respect to $\MB{f}^{-k}$ in the space $\MC{V}_{-}$ and we denote
$$
\MB{\Phi}={\tsum_{r_1,\ldots,r_M,s_1,\ldots,s_M}}
\MB{f}^{r_1}\otimes\ldots\otimes\MB{f}^{r_M}\otimes
\MB{e}_{-s_1}\otimes\ldots\otimes\MB{e}_{-s_M}\otimes
\Phi^{s_1,\ldots,s_M}_{r_1,\ldots,r_M}
$$
we can write the general shape of Bethe vectors in the
form\footnote{Another option is to introduce
$\tilde{\MB{B}}(u)={\tsum_{i,k=1}^n}\MB{E}^{-k}_i\otimes
T^i_{-k}(u)$
$$
\begin{array}{l}
\tilde{\MB{B}}_{1,\ldots,M}(\vec{u})=
\tilde{\MB{B}}_1(u_1)\otimes\tilde{\MB{B}}_2(u_2)\otimes\ldots\otimes\tilde{\MB{B}}_M(u_M)\\[6pt]
\tilde{\MB{\Phi}}={\tsum_{r_1,\ldots,r_M,s_1,\ldots,s_M=1}^n}
\MB{E}^{r_1}_{-s_1}\otimes\MB{E}^{r_2}_{-s_2}\otimes\ldots\otimes\MB{E}^{r_M}_{-s_M}\otimes
\Phi^{s_1,\ldots,s_M}_{r_1,\ldots,r_M}
\end{array}
$$
and write the Bethe vectors in the form
$\MF{V}(\vec{u})=
\Tr_{1,\ldots,M}\bigl(\tilde{\MB{B}}_{1,\ldots,M}(\vec{u})\tilde{\MB{\Phi}}\bigr)$}
$$
\MF{V}(\vec{u})= \Bigl<\MB{B}_{1,\ldots,M}(\vec{u}),\MB{\Phi}\Bigr>.
$$

\section{Commutation relations $\MB{T}_0^{(\pm)}(x)\MB{B}_{1,\ldots,M}(\vec{u})$}
 \label{sec:5}

On the space
$\MC{V}_0\otimes\MC{V}_{1_+}^*\otimes\MC{V}_{1_-}\otimes\MC{A}$ we
define\footnote{Further we will use the Yang--Baxter equation on
different spaces. Therefore, we give their equivalent shapes. If we
write the R--matrix in the form
$$
\wh{\MB{R}}_{1,2}(x)={\tsum_{i,k,r,s}}R^{k,s}_{i,r}(x)\MB{E}^i_k\otimes\MB{E}^r_s
$$
the Yang--Baxter equation
$$
\wh{\MB{R}}_{1,2}(x)\wh{\MB{R}}_{1,3}(xy)\wh{\MB{R}}_{2,3}(y)=
\wh{\MB{R}}_{2,3}(y)\wh{\MB{R}}_{1,3}(xy)\wh{\MB{R}}_{1,2}(x)
$$
gives the relations
\begin{equation}
 \label{YB-sou}
{\tsum_{a,b,c}}R^{k,s}_{a,b}(x)R^{a,q}_{i,c}(xy)R^{b,c}_{r,p}(y)=
{\tsum_{a,b,c}}R^{a,b}_{i,r}(x)R^{k,c}_{a,p}(xy)R^{s,q}_{b,c}(y)
\end{equation}
If we define
$$
\wh{\MB{R}}_{1,2^*}={\tsum_{i,k,r,s}}R^{k,s}_{i,r}(x)\MB{E}^i_k\otimes\MB{F}^r_s\,,\hskip10mm
\wh{\MB{R}}_{1^*,2^*}(x)={\tsum_{i,k,r,s}}R^{k,s}_{i,r}(x)\MB{F}^i_k\otimes\MB{F}^r_s\,,
$$
it is possible to write equality (\ref{YB-sou}) in one of the
following shapes:
$$
\begin{array}{l}
\MB{R}_{1,2}(x)\wh{\MB{R}}_{2,3^*}(y)\wh{\MB{R}}_{1,3^*}(xy)=
\wh{\MB{R}}_{1,3^*}(xy)\wh{\MB{R}}_{2,3^*}(y)\MB{R}_{1,2}(x)\\[4pt]
\wh{\MB{R}}_{1,3^*}(xy)\wh{\MB{R}}_{1,2^*}(x)\wh{\MB{R}}_{2^*,3^*}(y)=
\wh{\MB{R}}_{2^*,3^*}(y)\wh{\MB{R}}_{1,2^*}(x)\wh{\MB{R}}_{1,3^*}(xy)\\[4pt]
\wh{\MB{R}}_{1^*,2^*}(x)\wh{\MB{R}}_{1^*,3^*}(xy)\wh{\MB{R}}_{2^*,3^*}(y)=
\wh{\MB{R}}_{2^*,3^*}(y)\wh{\MB{R}}_{1^*,3^*}(xy)\wh{\MB{R}}_{1^*,2^*}(x)
\end{array}
$$
}%
$$
\begin{array}{l}
\wh{\MB{T}}^{(+)}_{0;1}(x;u)=
\bigl(\wh{\MB{R}}^{(+,+)}_{0,1^*}(xu^{-1})\bigr)^{-1}
\MB{T}^{(+)}_0(x)\wh{\MB{R}}^{(+,-)}_{0,1}(xu^{-1})\\[6pt]
\wh{\MB{T}}^{(-)}_{0;1}(x;u)=
\bigl(\wh{\MB{R}}^{(-,+)}_{0,1^*}(xu^{-1})\bigr)^{-1}
\MB{T}^{(-)}_0(x)\wh{\MB{R}}^{(-,-)}_{0,1}(xu^{-1})
\end{array}
$$
where
$$
\begin{array}{l}
\bigl(\wh{\MB{R}}^{(+,+)}_{0,1^*}(x)\bigr)^{-1}=
\dfrac1{f(x^{-1})}\Bigl({\tsum_{i,k=1;\,i\neq k}^n}
\MB{E}^i_i\otimes\MB{F}^k_k\otimes\MB{I}_-+
f(x^{-1}){\tsum_{i=1}^n}\MB{E}^i_i\otimes\MB{F}^i_i\otimes\MB{I}_-+\\[9pt]
\hskip20mm+
g(x^{-1}){\tsum_{1\leq i<k\leq n}}
\MB{E}^i_k\otimes\MB{F}^k_i\otimes\MB{I}_--
g(x){\tsum_{1\leq k<i\leq n}}
\MB{E}^i_k\otimes\MB{F}^k_i\otimes\MB{I}_-\Bigr)\\[12pt]
\bigl(\wh{\MB{R}}^{(-,+)}_{0,1^*}(x)\bigr)^{-1}=
{\tsum_{i,k=1;\,i\neq k}^n}
\MB{E}^{-i}_{-i}\otimes\MB{F}^k_k\otimes\MB{I}_-+
f(xq){\tsum_{i=1}^n}
\MB{E}^{-i}_{-i}\otimes\MB{F}^i_i\otimes\MB{I}_-+\\[9pt]
\hskip20mm+
g(xq){\tsum_{1\leq i<k\leq n}}q^{k-i}
\MB{E}^{-i}_{-k}\otimes\MB{F}^i_k\otimes\MB{I}_--
g(x^{-1}q^{-1}){\tsum_{1\leq k<i\leq n}}q^{k-i}
\MB{E}^{-i}_{-k}\otimes\MB{F}^i_k\otimes\MB{I}_-\\[12pt]
\wh{\MB{R}}^{(+,-)}_{0,1}(x)=
{\tsum_{i,k=1;\,i\neq k}^n}
\MB{E}^i_i\otimes\MB{I}^*_+\otimes\MB{E}^{-k}_{-k}+
f(x^{-1}q){\tsum_{i=1}^n}
\MB{E}^i_i\otimes\MB{I}^*_+\otimes\MB{E}^{-i}_{-i}+\\[9pt]
\hskip20mm+
g(x^{-1}q){\tsum_{1\leq i<k\leq n}}q^{k-i}
\MB{E}^i_k\otimes\MB{I}^*_+\otimes\MB{E}^{-i}_{-k}-
g(xq^{-1}){\tsum_{1\leq k<i\leq n}}q^{k-i}
\MB{E}^i_k\otimes\MB{I}^*_+\otimes\MB{E}^{-i}_{-k}\\[12pt]
\wh{\MB{R}}^{(-,-)}_{0,1}(x)=
\dfrac1{f(x)}\Bigl({\tsum_{i,k=1;\,i\neq k}^n}
\MB{E}^{-i}_{-i}\otimes\MB{I}^*_+\otimes\MB{E}^{-k}_{-k}+
f(x){\tsum_{i=1}^n}
\MB{E}^{-i}_{-i}\otimes\MB{I}^*_+\otimes\MB{E}^{-i}_{-i}+\\[9pt]
\hskip20mm+
g(x){\tsum_{1\leq i<k\leq n}}
\MB{E}^{-i}_{-k}\otimes\MB{I}^*_+\otimes\MB{E}^{-k}_{-i}-
g(x^{-1}){\tsum_{1\leq k<i\leq n}}
\MB{E}^{-i}_{-k}\otimes\MB{I}^*_+\otimes\MB{E}^{-k}_{-i}\Bigr)
\end{array}
$$

\underbar{\textbf{Lemma 3.}} In the RTT--algebra of
$U_q\bigl(\MR{sp}(2n)\bigr)$ type the relations
$$
\begin{array}{l}
\MB{T}^{(+)}_0(x)\Bigl<\MB{B}_1(u),\MB{f}^r\otimes\MB{e}_{-s}\Bigr>=
f(x^{-1}u)\Bigl<\MB{B}_1(u),\wh{\MB{T}}^{(+)}_{0;1}(x;u)
\bigl(\MB{I}\otimes\MB{f}^r\otimes\MB{e}_{-s}\bigr)\Bigr>+\\[6pt]
\hskip30mm+
g(xu^{-1})\Bigl<\MB{B}_1(x),\wh{\MB{T}}^{(+)}_{0;1}(u;u)
\bigl(\MB{I}\otimes\MB{f}^r\otimes\MB{e}_{-s}\bigr)\Bigr>\\[12pt]
\MB{T}^{(-)}_0(x)\Bigl<\MB{B}_1(u),\MB{f}^r\otimes\MB{e}_{-s}\Bigr>=
f(xu^{-1})\Bigl<\MB{B}_1(u),\wh{\MB{T}}^{(-)}_{0;1}(x;u)
\bigl(\MB{I}\otimes\MB{f}^r\otimes\MB{e}_{-s}\bigr)\Bigr>-\\[6pt]
\hskip30mm-
g(xu^{-1}))\Bigl<\MB{B}_1(x),\wh{\MB{T}}^{(-)}_{0;1}(u;u)
\bigl(\MB{I}\otimes\MB{f}^r\otimes\MB{e}_{-s}\bigr)\Bigr>
\end{array}
$$
are valid.

\bigskip

For ordered $M$--tuples $\vec{u}=(u_1,\ldots,u_M)$, let $\ol{u}$
denote the set $\ol{u}=\{u_1,\ldots,u_M\}$. We define
$$
\begin{array}{ll}
\vec{u}_k=(u_1,\ldots,u_{k-1},u_{k+1},\ldots,u_M)\,,\qquad&
\ol{u}_k=\ol{u}\setminus\{u_k\}=\{u_1,\ldots,u_{k-1};u_{k+1},\ldots,u_M\}\,,\\[4pt]
F(x;\ol{u}^{-1})={\tprod_{k=1}^M}f(xu_k^{-1})\,,\qquad&
F(x^{-1},\ol{u})={\tprod_{k=1}^M}f(x^{-1}u_k)\,.
\end{array}
$$
and introduce the operators
$$
\begin{array}{l}
\wh{\MB{T}}^{(+)}_{0;1,\ldots,M}(x;\vec{u})=
\bigl(\wh{\MB{R}}^{(+,+)}_{0,1^*}(xu_1^{-1})\bigr)^{-1}\ldots
\bigl(\wh{\MB{R}}^{(+,+)}_{0,M^*}(xu_M^{-1})\bigr)^{-1}
\MB{T}^{(+)}_0(x)\\[4pt]
\hskip80mm
\wh{\MB{R}}^{(+,-)}_{0,M}(xu_M^{-1})\ldots
\wh{\MB{R}}^{(+,-)}_{0,1}(xu_1^{-1})\\[9pt]
\wh{\MB{T}}^{(-)}_{0;1,\ldots,M}(x;\vec{u})=
\bigl(\wh{\MB{R}}^{(-,+)}_{0,1^*}(xu_1^{-1})\bigr)^{-1}\ldots
\bigl(\wh{\MB{R}}^{(-,+)}_{0,M^*}(xu_M^{-1})\bigr)^{-1}
\MB{T}^{(-)}_0(x)\\[4pt]
\hskip80mm
\wh{\MB{R}}^{(-,-)}_{0,M}(xu_M^{-1})\ldots
\wh{\MB{R}}^{(-,-)}_{0,1}(xu_1^{-1})\\[9pt]
\MB{B}_{k;1,\ldots,M}(x;\vec{u}_k)=
\MB{B}_k(x)\otimes\MB{B}_1(u_1)\otimes\ldots\otimes\MB{B}_{k-1}(u_{k-1})\otimes
\MB{B}_{k+1}(u_{k+1})\otimes\ldots\otimes\MB{B}_M(u_M)
\end{array}
$$

\underbar{\textbf{Proposition  2.}} The following relationships are
applied:
$$
\begin{array}{l}
\MB{T}^{(+)}_0(x)\Bigl<\MB{B}_{1,\ldots,M}(\vec{u}),\MB{\Phi}\Bigr>=
F(x^{-1};\ol{u})\Bigl<\MB{B}_{1,\ldots,M}(\vec{u}),
\wh{\MB{T}}^{(+)}_{0;1,\ldots,M}(x;\vec{u})\MB{\Phi}\Bigr>+\\[6pt]
\hskip5mm+
{\tsum_{u_k\in\ol{u}}}g(xu_k^{-1})F(u_k^{-1};\ol{u}_k)
\Bigl<\MB{B}_{k;1,\ldots,M}(x;\vec{u}_k),
\bigl(\wh{\MB{R}}^{(+,+)}_{1^*,\ldots,k^*}(\vec{u})\bigr)^{-1}
\wh{\MB{R}}^{(-,-)}_{1,\ldots,k}(\vec{u})
\wh{\MB{T}}^{(+)}_{0;1,\ldots,M}(u_k;\vec{u})\MB{\Phi}\Bigr>\\[12pt]
\MB{T}^{(-)}_0(x)\Bigl<\MB{B}_{1,\ldots,M}(\vec{u}),\MB{\Phi}\Bigr>=
F(x;\ol{u}^{-1})\Bigl<\MB{B}_{1,\ldots,M}(\vec{u}),
\wh{\MB{T}}^{(-)}_{0;1,\ldots,M}(x;\vec{u})\MB{\Phi}\Bigr>-\\[6pt]
\hskip5mm-
{\tsum_{u_k\in\ol{u}}}g(xu_k^{-1})
F(u_k;\ol{u}_k^{-1}\Bigl<\MB{B}_{k;1,\ldots,M}(x;\vec{u}_k),
\bigl(\wh{\MB{R}}^{(+,+)}_{1^*,\ldots,k^*}(\vec{u})\bigr)^{-1}
\wh{\MB{R}}^{(-,-)}_{1,\ldots,k}(\vec{u})
\wh{\MB{T}}^{(-)}_{0;1,\ldots,M}(u_k;\vec{u})\MB{\Phi}\Bigr>
\end{array}
$$
where
$$
\begin{array}{l}
\wh{\MB{R}}^{(+,+)}_{1^*,\ldots,k^*}(\vec{u})=
\wh{\MB{R}}^{(+,+)}_{(k-1)^*,k^*}(u_{k-1}u_k^{-1})\ldots
\wh{\MB{R}}^{(+,+)}_{2^*,k^*}(u_2u_k^{-1})
\wh{\MB{R}}^{(+,+)}_{1^*,k^*}(u_1u_k^{-1})\\[6pt]
\wh{\MB{R}}^{(-,-)}_{1,\ldots,k}(\vec{u})=
\wh{\MB{R}}^{(-,-)}_{1,k}(u_1u_k^{-1})\wh{\MB{R}}^{(-,-)}_{2,k}(u_2u_k^{-1})\ldots
\wh{\MB{R}}^{(-,-)}_{k-1,k}(u_{k-1}u_k^{-1})\\[9pt]
\wh{\MB{R}}^{(+,+)}_{1^*,2^*}(x)=
\dfrac1{f(x)}\Bigl({\tsum_{i,k=1;\,i\neq k}^n}\MB{F}^i_i\otimes\MB{F}^k_k+
f(x){\tsum_{i=1}^n}\MB{F}^i_i\otimes\MB{F}^i_i-\\[9pt]
\hskip30mm-
g(x^{-1}){\tsum_{1\leq i<k\leq n}}\MB{F}^i_k\otimes\MB{F}^k_i+
g(x){\tsum_{1\leq k<i\leq n}}\MB{F}^i_k\otimes\MB{F}^k_i\Bigr)
\end{array}
$$

\section{Bethe conditions and eigenvectors of the operator $H(x)$}
 \label{sec:6}

Let us denote by $\wh{T}^i_k(x;\vec{u})$ and
$\wh{T}^{-i}_{-k}(x;\vec{u})$ the operators defined by the relations
$$
\wh{\MB{T}}^{(+)}_{0;1,\ldots,M}(x;\vec{u})=
{\tsum_{i,k=1}^n}\MB{E}^k_i\otimes\wh{T}^i_k(x;\vec{u})\,,
\hskip15mm \wh{\MB{T}}^{(-)}_{0;1,\ldots,M}(x;\vec{u})=
{\tsum_{i,k=1}^n}\MB{E}^{-k}_{-i}\otimes\wh{T}^{-i}_{-k}(x;\vec{u})\,.
$$
The following statement, which gives part of the Bethe conditions,
follows from the previous part.

\medskip

\underbar{\textbf{Theorem 1.}} Let $\MB{\Phi}$ be common eigenvector
of the operators
$$
\begin{array}{l}
\wh{H}^{(+)}_{1,\ldots,M}(x;\vec{u})=
\Tr_0\Bigl(\wh{\MB{T}}^{(+)}_{0;1,\ldots,M}(x;\vec{u})\Bigr)=
{\tsum_{i=1}^n}\wh{T}^i_i(x;\vec{u})\,,\\[6pt]
\wh{H}^{(-)}_{1,\ldots,M}(x;\vec{u})=
\Tr_0\Bigl(\wh{\MB{T}}^{(-)}_{0;1,\ldots,M}(x;\vec{u})\Bigr)=
{\tsum_{i=1}^n}\wh{T}^{-i}_{-i}(x;\vec{u})
\end{array}
$$
with eigenvalues $\wh{E}^{(+)}_{1,\ldots,M}(x;\vec{u})$ and
$\wh{E}^{(-)}_{1,\ldots,M}(x;\vec{u})$. If for each $u_k\in\ol{u}$
the relations
\begin{equation}
 \label{BP-1}
\wh{E}^{(+)}_{1,\ldots,M}(u_k;\vec{u})F(u_k^{-1};\ol{u}_k)=
\wh{E}^{(-)}_{1,\ldots,M}(u_k;\vec{u})F(u_k;\ol{u}_k^{-1})
\end{equation}
are true, then $\Bigl<\MB{B}_{1,\ldots,M}(\vec{u}),\MB{\Phi}\Bigr>$
is the eigenvector of the operator $H(x)=H^{(+)}(x)+H^{(-)}(x)$,
where $H^{(\pm)}(x)=\Tr\bigl(\MB{T}^{(\pm)}_0(x)\bigr)$ with the
eigenvalue
$$
E_{1,\ldots,M}(x;\vec{u})=
\wh{E}^{(+)}_{1,\ldots,M}(x;\vec{u})F(x^{-1};\ol{u})+
\wh{E}^{(-)}_{1,\ldots,M}(x;\vec{u})F(x;\ol{u}^{-1})\,.
$$

\bigskip

Thus, to find the eigenvectors of the operators $H(x)$, it is
sufficient to find common eigenvectors of the operators
$\wh{H}^{(+)}_{1,\ldots,M}(x;\vec{u})$ and
$\wh{H}^{(-)}_{1,\ldots,M}(x;\vec{u})$.

\medskip

\underbar{\textbf{Theorem 2.}} The operators
$\wh{\MB{T}}^{(\pm)}_{0;1,\ldots,M}(x;\vec{u})$ fulfill the
RTT--equation\footnote{We cannot write this RTT--equation, even for
$M=1$, using the matrix $\tilde{\MB{B}}(u)$. Therefore, we prefer to
use the formulation using the spaces $\MC{V}_{\pm}$ and
$\MC{V}^*_{\pm}$.}
$$
\MB{R}^{(\epsilon,\epsilon')}_{0,0'}(xy^{-1})
\wh{\MB{T}}^{(\epsilon)}_{0;1,\ldots,M}(x;\vec{u})
\wh{\MB{T}}^{(\epsilon')}_{0';1,\ldots,M}(y;\vec{u})=
\wh{\MB{T}}^{(\epsilon')}_{0';1,\ldots,M}(y;\vec{u})
\wh{\MB{T}}^{(\epsilon)}_{0;1,\ldots,M}(x;\vec{u})
\MB{R}^{(\epsilon,\epsilon')}_{0,0'}(xy^{-1})
$$
for any $\vec{u}$ and $\epsilon,\,\epsilon'=\pm$ and thus generate
the RTT--algebra $\tilde{A}_n$.

\medskip

\underbar{\textbf{Theorem 3.}} The vector
$$
\wh{\MB{\Omega}}=
\underbrace{\MB{f}^1\otimes\ldots\otimes\MB{f}^1}_{M\times}\otimes
\underbrace{\MB{e}_{-1}\otimes\ldots\otimes\MB{e}_{-1}}_{M\times}\otimes
\omega
$$
is a vacuum vector for representation of the RTT--algebra
$\tilde{\MC{A}}_n$ with the weights
$$
\begin{array}{lll}
\mu_1(x;\vec{u})=\lambda_1(x)F(x^{-1}q;\ol{u})\qquad&
\mu_k(x;\vec{u})=\lambda_k(x)F(xq^{-1};\ol{u}^{-1})\quad&
k=2,\,\ldots,\,n\\[9pt]
\mu_{-1}(x;\vec{u})=\lambda_{-1}(x)F(xq;\ol{u}^{-1})\qquad&
\mu_{-k}(x;\vec{u})=
\lambda_{-k}(x)F(x^{-1}q^{-1};\ol{u})\quad&
k=2,\,\ldots,\,n\,.
\end{array}
$$

\bigskip

So to find eigenvectors of the operators $H(x)$ for the RTT--algebra
of $U_q\bigl(\MR{sp}(2n)\bigr)$ type, it is enough to formulate the
Bethe ansatz for the RTT--algebra $\tilde{\MC{A}}_n$.

\section*{Appendix}
 \label{sec:A}

\subsection*{A1 Commutation relations in the RTT--algebra of
$U_q\bigl(\MR{sp}(2n)\bigr)$ type}

If we denote
$$
\theta_{r-i}=\left\{\begin{array}{lcl}
1&\quad\MR{for}\quad&r>i\\
0&\quad\MR{for}\quad&r\leq i\end{array}\right.
$$
we obtain  from the RTT-equation
$$
\MB{R}_{1,2}(xy^{-1})\MB{T}_1(x)\MB{T}_2(y)=
\MB{T}_2(y)\MB{T}_1(x)\MB{R}_{1,2}(xy^{-1})
$$
that the generators $T^i_k(x)$ fulfill  these commutation relations
\begin{equation}
 \label{RTT-1}
\begin{array}{l}
T^i_k(x)T^r_s(y)+
\delta_{i,r}\bigl(f(xy^{-1})-1\bigr)T^i_k(x)T^r_s(y)+
\delta_{i,-r}\bigl(f(x^{-1}yq^{-n-1})-1\bigr)T^i_k(x)T^r_s(y)+\\[9pt]
\hskip20mm+ g(xy^{-1})\theta_{r-i}T^r_k(x)T^i_s(y)-
g(x^{-1}y)\theta_{i-r}T^r_k(x)T^i_s(y)-\\[9pt]
\hskip30mm-
\delta_{i,-r}g(xy^{-1}q^{n+1}){\tsum_{p>i}}q^{i-p}\epsilon_i\epsilon_p
T^p_k(x)T^{-p}_s(y)+\\[9pt]
\hskip40mm+
\delta_{i,-r}g(x^{-1}yq^{-n-1}){\tsum_{p<i}}q^{i-p}\epsilon_i\epsilon_p
T^p_k(x)T^{-p}_s(y)=\\[12pt]
= T^r_s(y)T^i_k(x)+
\delta_{k,s}\bigl(fxy^{-1})-1\bigr)T^r_s(y)T^i_k(x)+
\delta_{k,-s}\bigl(f(x^{-1}yq^{-n-1})-1\bigr)T^r_s(y)T^i_k(x)+\\[9pt]
\hskip20mm+ g(xy^{-1})\theta_{k-s}T^r_k(y)T^i_s(x)-
g(x^{-1}y)\theta_{s-k}T^r_k(y)T^i_s(x)-\\[9pt]
\hskip30mm-
\delta_{k,-s}g(xy^{-1}q^{n+1}){\tsum_{p<k}}q^{p-k}\epsilon_k\epsilon_p
T^r_{-p}(y)T^i_p(x)+\\[9pt]
\hskip40mm+
\delta_{k,-s}g(x^{-1}yq^{-n-1}){\tsum_{p>k}}q^{p-k}\epsilon_k\epsilon_p
T^r_{-p}(y)T^i_p(x)
\end{array}
\end{equation}
and further the RTT--equation
$$
\MB{T}_1(x)\MB{T}_2(y)\MB{R}^{-1}_{1,2}(xy^{-1})=
\MB{R}^{-1}_{1,2}(xy^{-1})\MB{T}_2(y)\MB{T}_1(x)\,.
$$
leads to the commutation relations
\begin{equation}
 \label{RTT-2}
\begin{array}{l}
T^i_k(x)T^r_s(y)+
\delta_{k,s}\bigl(f(x^{-1}y)-1\bigr)T^i_k(x)T^r_s(y)+
\delta_{k,-s}\bigl(f(xy^{-1}q^{-n-1})-1\bigr)T^i_k(x)T^r_s(y)-\\[9pt]
\hskip20mm- g(xy^{-1})\theta_{k-s}T^i_s(x)T^r_k(y)+
g(x^{-1}y)\theta_{s-k}T^i_s(x)T^r_k(y)+\\[9pt]
\hskip30mm+
\delta_{k,-s}g(xy^{-1}q^{-n-1}){\tsum_{p<k}}q^{k-p}\epsilon_k\epsilon_p
T^i_p(x)T^r_{-p}(y)-\\[9pt]
\hskip40mm-
\delta_{k,-s}g(x^{-1}yq^{n+1}){\tsum_{p>k}}q^{k-p}\epsilon_k\epsilon_p
T^i_p(x)T^r_{-p}(y)=\\[12pt]
= T^r_s(y)T^i_k(x)+
\delta_{i,r}\bigl(f(x^{-1}y)-1\bigr)T^r_s(y)T^i_k(x)+
\delta_{i,-r}\bigl(f(xy^{-1}q^{-n-1})-1\bigr)
T^r_s(y)T^i_k(x)-\\[9pt]
\hskip20mm- g(xy^{-1})\theta_{r-i}T^i_s(y)T^r_k(x)+
g(x^{-1}y)\theta_{i-r}T^i_s(y)T^r_k(x)+\\[9pt]
\hskip30mm+
\delta_{i,-r}g(xy^{-1}q^{-n-1}){\tsum_{p>i}}q^{p-i}\epsilon_i\epsilon_p
T^{-p}_s(y)T^p_k(x)-\\[9pt]
\hskip40mm-
\delta_{i,-r}g(x^{-1}yq^{n+1}){\tsum_{p<i}}q^{p-i}\epsilon_i\epsilon_p
T^{-p}_s(y)T^p_k(x)
\end{array}
\end{equation}

\subsection*{A2\quad Proof of Lemma 1}

It is enough to prove that if $T^{-i}_k(x)\Omega=0$ is valid  for
$\Omega\in\MC{W}_0$, the relations
$$
T^{-i}_k(x)T^r_s(y)\Omega=0\,,\qquad
T^{-i}_k(x)T^{-r}_{-s}(y)\Omega=0
$$
also hold true.

The commutation relations (\ref{RTT-1}) give
$$
\begin{array}{l}
T^{-i}_k(x)T^r_s(y)\Omega+
\delta_{i,r}\Bigl(\bigl(f(x^{-1}yq^{-n-1})-1\bigr)T^{-i}_k(x)T^r_s(y)\Omega-\\[6pt]
\hskip5mm-
g(xy^{-1}q^{n+1}){\tsum_{p=1}^{i-1}}q^{p-i}T^{-p}_k(x)T^p_s(y)\Omega+
g(x^{-1}yq^{-n-1}){\tsum_{p=i+1}^n}q^{p-i}T^{-p}_k(x)T^p_s(y)\Omega\Bigr)=0
\end{array}
$$
and relations  (\ref{RTT-2}) lead to the equality
$$
\begin{array}{l}
T^{-i}_k(x)T^{-r}_{-s}(y)\Omega+
\delta_{k,s}\Bigl(\bigl(f(xy^{-1}q^{-n-1})-1\bigr)T^{-i}_k(x)T^{-r}_{-s}(y)\Omega+\\[6pt]
\hskip5mm+
g(xy^{-1}q^{-n-1}){\tsum_{p=1}^{k-1}}q^{k-p}T^{-i}_p(x)T^{-r}_{-p}(y)\Omega-
g(x^{-1}yq^{n+1}){\tsum_{p=k+1}^n}q^{k-p}T^{-i}_p(x)T^{-r}_{-p}(y)\Omega\Bigr)=0
\end{array}
$$
So the first relation holds for $i\neq r$ and the second relation for $k\neq s$.

It remains to show that
\begin{equation}
 \label{Lemma-1}
\begin{array}{l}
f(x^{-1}yq^{-n-1})Z_i- g(xy^{-1}q^{n+1}){\tsum_{p=1}^{i-1}}Z_p+
g(x^{-1}yq^{-n-1}){\tsum_{p=i+1}^n}Z_p=0\\[6pt]
f(xy^{-1}q^{-n-1})\wh{Z}_k+
g(xy^{-1}q^{-n-1}){\tsum_{p=1}^{k-1}}\wh{Z}_p-
g(x^{-1}yq^{n+1}){\tsum_{p=k+1}^n}\wh{Z}_p=0\,,
\end{array}
\end{equation}
where
$$
Z_i=q^iT^{-i}_k(x)T^i_s(y)\Omega\,,\qquad
\wh{Z}_k=q^{-k}T^{-i}_k(x)T^{-r}_{-k}(y)\Omega
$$
is true. If we subtract the first two relationships for $i+1$ and
$i$ and the second for $k+1$ and $k$, we get
$$
\begin{array}{l}
\Bigl(f(x^{-1}yq^{-n-1})-g(x^{-1}yq^{-n-1})\Bigr)Z_{i+1}=
\Bigl(f(x^{-1}yq^{-n-1})+g(xy^{-1}q^{n+1})\Bigr)Z_i\\[4pt]
\Bigl(f(xy^{-1}q^{-n-1})+g(x^{-1}yq^{n+1})\Bigr)\wh{Z}_{k+1}=
\Bigl(f(xy^{-1}q^{-n-1})-g(xy^{-1}q^{-n-1})\Bigr)\wh{Z}_k
\end{array}
$$
or
$$
q^{-1}Z_{i+1}=qZ_i\,,\qquad q\wh{Z}_{k+1}=q^{-1}\wh{Z}_k
$$
Thus, $Z_i=cq^{2i}$ and $\wh{Z}_k=\wh{c}q^{-2k}$, where $c$ and
$\wh{c}$ are some constants. Substituting the latter into
(\ref{Lemma-1}), we obtain $c=\wh{c}=0$.

\subsection*{A3\quad Proof of Lemma 2}

It is easy to see that the RTT--algebras $\MC{A}^{(\pm)}$ are the
RTT--subalgebras of the RTT--algebra of $U_q\bigl(\MR{sp}(2n)\bigr)$
type. Therefore, the relations
$$
\begin{array}{l}
\MB{R}^{(+,+)}_{1,2}(xy^{-1})
\MB{T}^{(+)}_1(x)\MB{T}^{(+)}_2(y)=
\MB{T}^{(+)}_2(y)\MB{T}^{(+_1)}_1(x)
\MB{R}^{(+,+)}_{1,2}(xy^{-1})\\[9pt]
\MB{R}^{(-,-)}_{1,2}(xy^{-1})
\MB{T}^{(-)}_1(x)\MB{T}^{(-)}_2(y)=
\MB{T}^{(-)}_2(y)\MB{T}^{(-)}_1(x)
\MB{R}^{(-,-)}_{1,2}(xy^{-1})
\end{array}
$$
are the RTT--relation
$$
\MB{R}_{1,2}(xy^{-1})\MB{T}_1(x)\MB{T}_2(y)=
\MB{T}_2(y)\MB{T}_1(x)\MB{R}_{1,2}(xy^{-1})
$$
restricted on the spaces $\MC{A}^{(+)}$ and $\MC{A}^{(-)}$.

Since $T^{-i}_k(x)=0$ on the space $\MC{W}_0$, we get using the
commutation relation (\ref{RTT-1}) that on the space $\MC{W}_0$
$$
\begin{array}{l}
T^{-i}_{-k}(x)T^r_s(y)+
\delta_{i,r}\bigl(f(x^{-1}yq^{-n-1})-1\bigr)T^{-i}_{-k}(x)T^r_s(y)-\\[6pt]
\hskip7mm- \delta_{i,r}g(xy^{-1}q^{n+1}){\tsum_{p=1}^{i-1}}q^{p-i}
T^{-p}_{-k}(x)T^p_s(y)+
\delta_{i,r}g(x^{-1}yq^{-n-1}){\tsum_{p=i+1}^n}q^{p-i}
T^{-p}_{-k}(x)T^p_s(y)=\\[9pt]
= T^r_s(y)T^{-i}_{-k}(x)+
\delta_{k,s}\bigl(f(x^{-1}yq^{-n-1})-1\bigr)T^r_s(y)T^{-i}_{-k}(x)-\\[6pt]
\hskip7mm- \delta_{k,s}g(xy^{-1}q^{n+1}){\tsum_{p=k+1}^n}q^{-p+k}
T^r_p(y)T^{-i}_{-p}(x)+
\delta_{k,s}g(x^{-1}yq^{-n-1}){\tsum_{p=1}^{k-1}}q^{-p+k}
T^r_p(y)T^{-i}_{-p}(x)
\end{array}
$$
hold true. The equation
$$
\MB{R}^{(-,+)}_{1,2}(xy^{-1})\MB{T}^{(-)}_1(x)\MB{T}^{(+)}_2(y)=
\MB{T}^{(+)}_2(y)\MB{T}^{(-)}_1(x)\MB{R}^{(-,+)}_{1,2}(xy^{-1})
$$
is only the matrix representation of these relations.

By narrowing commutation relations (\ref{RTT-2}) to the space
$\MC{W}_0$, we have
$$
\begin{array}{l}
T^i_k(x)T^{-r}_{-s}(y)+
\delta_{k,s}\bigl(f(xy^{-1}q^{-n-1})-1\bigr)T^i_k(x)T^{-r}_{-s}(y)+\\[6pt]
\hskip10mm+
\delta_{k,s}g(xy^{-1}q^{-n-1}){\tsum_{p=1}^{k-1}}q^{k-p}
T^i_p(x)T^{-r}_{-p}(y)-
\delta_{k,s}g(x^{-1}yq^{n+1}){\tsum_{p=k+1}^n}q^{k-p}
T^i_p(x)T^{-r}_{-p}(y)=\\[9pt]
= T^{-r}_{-s}(y)T^i_k(x)+
\delta_{i,r}\bigl(f(xy^{-1}q^{-n-1})-1\bigr)
T^{-r}_{-s}(y)T^i_k(x)+\\[6pt]
\hskip10mm+
\delta_{i,r}g(xy^{-1}q^{-n-1}){\tsum_{p=i+1}^n}q^{p-i}T^{-p}_{-s}(y)T^p_k(x)-
\delta_{i,r}g(x^{-1}yq^{n+1}){\tsum_{p=1}^{i-1}}q^{p-i}T^{-p}_{-s}(y)T^p_k(x)
\end{array}
$$
These relations can be rewritten as
$$
\MB{T}^{(+)}_1(x)\MB{T}^{(-)}_2(y)
\bigl(\MB{R}^{(+,-)}_{1,2}(xy^{-1}\bigr)^{-1}=
\bigl(\MB{R}^{(+,-)}_{1,2}(xy^{-1}\bigr)^{-1}
\MB{T}^{(-)}_2(y)\MB{T}^{(+)}_1(x)
$$
where
$$
\begin{array}{l}
\bigl(\MB{R}^{(+,-)}_{1,2}(x)\bigr)^{-1}=
{\tsum_{i,k=1;\,i\neq k}^n}\MB{E}^i_i\otimes\MB{E}^{-k}_{-k}+
f(xq^{-n-1}){\tsum_{i=1}^n}\MB{E}^i_i\otimes\MB{E}^{-i}_{-i}+\\[9pt]
\hskip15mm+ g(xq^{-n-1}){\tsum_{1\leq k<i\leq
n}}q^{i-k}\MB{E}^i_k\otimes\MB{E}^{-i}_{-k}- g(x^{-1}q^{n+1})
{\tsum_{1\leq i<k\leq n}}q^{i-k}\MB{E}^i_k\otimes\MB{E}^{-i}_{-k}
\end{array}
$$
The proven equality
$$
\MB{R}^{(+,-)}_{1,2}(xy^{-1})\MB{T}^{(+)}_1(x)\MB{T}^{(-)}_2(y)=
\MB{T}^{(-)}_2(y)\MB{T}^{(+)}_1(x)\MB{R}^{(+,-)}_{1,2}(xy^{-1})
$$
then results from the relations
$$
\MB{R}^{(+,-)}_{1,2}(x)\bigl(\MB{R}^{(+,-)}_{1,2}(x)\bigr)^{-1}=
\bigl(\MB{R}^{(+,-)}_{1,2}(x)\bigr)^{-1}\MB{R}^{(+,-)}_{1,2}(x)=
\MB{I}^{(+)}\otimes\MB{I}^{(-)}
$$
where $\MB{I}^{(+)}={\tsum_{i=}^n}\MB{E}^i_i$ and
$\MB{I}^{(-)}={\tsum_{i=1}^n}E^{-i}_{-i}$, which can be checked by
direct calculations.

\subsection*{A4\quad Proof of Lemma 3}

These relationships are actually the commutation relations for
$T^i_k(x)T^r_{-s}(u)$ and $T^{-i}_{-k}(x)T^r_{-s}(u)$.

Using commutation relations (\ref{RTT-2}), we obtain
$$
\begin{array}{l}
T^i_k(x)T^r_{-s}(u)+
\delta_{k,s}\bigl(f(xu^{-1}q^{-n-1})-1\bigr)T^i_k(x)T^r_{-s}(u)+\\[6pt]
\hskip20mm+ \delta_{k,s}g(xu^{-1}q^{-n-1}){\tsum_{p=1}^{k-1}}q^{k-p}
T^i_p(x)T^r_{-p}(u)-\\[6pt]
\hskip40mm- \delta_{k,s}g(x^{-1}uq^{n+1}){\tsum_{p=k+1}^n}q^{k-p}
T^i_p(x)T^r_{-p}(u)=\\[9pt]
= T^r_{-s}(u)T^i_k(x)+
\delta_{i,r}\bigl(f(x^{-1}u)-1\bigr)T^r_{-s}(u)T^i_k(x)-\\[6pt]
\hskip20mm- g(xu^{-1})\theta_{r-i}T^i_{-s}(u)T^r_k(x)+
g(x^{-1}u)\theta_{i-r}T^i_{-s}(u)T^r_k(x)+\\[6pt]
\hskip40mm+ g(xu^{-1})T^i_{-s}(x)T^r_k(u)+
\delta_{k,s}g(xu^{-1}q^{-n-1}){\tsum_{p=1}^n}q^{k+p}
T^i_{-p}(x)T^r_p(u)\,.
\end{array}
$$
If we denote
$$
\tilde{\MB{B}}_1(u)={\tsum_{r,s=1}^n}\MB{E}^{-s}_r\otimes
T^r_{-s}(u)\,, \qquad
\MB{P}^{(+,+)}_{0,1}={\tsum_{i,k=1}^n}\MB{E}^i_k\otimes\MB{E}^k_i\,,
\qquad
\MB{Q}^{(+,-)}_{0,1}={\tsum_{i,k=1}^n}q^{i+k}\MB{E}^i_k\otimes\MB{E}^{-i}_{-k}\,,
$$
these relationships can be rewritten as
$$
\begin{array}{l}
\MB{T}^{(+)}_0(x)\tilde{\MB{B}}_1(u)\bigl(\MB{R}^{(+,-)}_{0,1}(xu^{-1})\bigr)^{-1}=
f(x^{-1}u)\bigl(\MB{R}^{(+,+)}_{0,1}(xu^{-1})\bigr)^{-1}
\tilde{\MB{B}}_1(u)\MB{T}^{(+)}_0(x)+\\[6pt]
\hskip20mm+
g(xu^{-1})\MB{P}^{(+,+)}_{0,1}\tilde{\MB{B}}_1(x)\MB{T}^{(+)}_0(u)
\Bigl(\MB{I}^{(+)}_0\otimes\MB{I}^{(-)}_1+
\dfrac{q^{-n-1}(xu^{-1}-x^{-1}u)}{xu^{-1}q^{-n-1}-x^{-1}uq^{n+1}}\,
\MB{Q}^{(+,-)}_{0,1}\Bigr)
\end{array}
$$
And since the relation
$$
\begin{array}{l}
\Bigl(\MB{I}^{(+)}_0\otimes\MB{I}^{(-)}_1+
\dfrac{q^{-n-1}(xu^{-1}-x^{-1}u)}{xu^{-1}q^{-n-1}-x^{-1}uq^{n+1}}\,
\MB{Q}^{(+,-)}_{0,1}\Bigr)\MB{R}^{(+,-)}_{0,1}(xu^{-1})=\\[9pt]
\hskip10mm= {\tsum_{i\neq k}}\MB{E}^i_i\otimes\MB{E}^{-k}_{-k}+
{\tsum_i} (q+q^{-1})\MB{E}^i_i\otimes\MB{E}^{-i}_{-i}+ {\tsum_{k<i}}
q^{k-i-1}\MB{E}^i_k\otimes\MB{E}^{-i}_{-k}+ {\tsum_{i<k}}q^{k-i+1}
\MB{E}^i_k\otimes\MB{E}^{-i}_{-k}
\end{array}
$$
holds, after multiplying the equality by the matrix
$\MB{R}^{(+,-)}_{0,1}(xu^{-1})$ from the right, we get the relation
$$
\begin{array}{l}
\MB{T}^{(+)}_0(x)\tilde{\MB{B}}_1(u)=
f(x^{-1}u)\bigl(\MB{R}^{(+,+)}_{0,1}(xu^{-1})\bigr)^{-1}
\tilde{\MB{B}}_1(u)\MB{T}^{(+)}_0(x)\MB{R}^{(+,-)}_{0,1}(xu^{-1})+\\[6pt]
\hskip20mm+
g(xu^{-1})\bigl(\MB{R}^{(+,+)}_{0,1}(1)\bigr)^{-1}\tilde{\MB{B}}_1(x)\MB{T}^{(+)}_0(u)
\MB{R}^{(+,-)}_{0,1}(1)
\end{array}
$$
where
$$
\begin{array}{l}
\MB{R}^{(+,+)}_{0,1}(1)=\bigl(\MB{R}^{(+,+)}_{0,1}(1)\bigr)^{-1}=\MB{P}^{(+,+)}_{0,1}\\[6pt]
\MB{R}^{(+,-)}_{0,1}(1)= {\tsum_{i,k=1,\,i\neq k}^n}
\MB{E}^i_i\otimes\MB{E}^{-k}_{-k}+ {\tsum_{i=1}^n}
(q+q^{-1})\MB{E}^i_i\otimes\MB{E}^{-i}_{-i}+\\[6pt]
\hskip40mm+ {\tsum_{1\leq k<i\leq n}}
q^{k-i-1}\MB{E}^i_k\otimes\MB{E}^{-i}_{-k}+ {\tsum_{1\leq i<k\leq
n}}q^{k-i+1} \MB{E}^i_k\otimes\MB{E}^{-i}_{-k}
\end{array}
$$
Similarly, using commutation relation (\ref{RTT-1}), it can be
proved that the relation
$$
\begin{array}{l}
\MB{T}^{(-)}_0(x)\tilde{\MB{B}}_1(u)=
f(xu^{-1})\bigl(\MB{R}^{(-,+)}_{0,1}(xu^{-1})\bigr)^{-1}
\tilde{\MB{B}}_1(u)\MB{T}^{(-)}_0(x)\MB{R}^{(-,-)}_{0,1}(xu^{-1})-\\[6pt]
\hskip30mm- g(xu^{-1}))\bigl(\MB{R}^{(-,+)}_{0,1}(1)\bigr)^{-1}
\tilde{\MB{B}}_1(x)\MB{T}^{(-)}_0(u)\MB{R}^{(-,-)}_{0,1}(1)
\end{array}
$$
is valid.

The relations from the Lemma are just a transcription of these
sessions into our formalism.

\subsection*{A5\quad Proof of Proposition 2\footnote{These claims can be
substantiated induction by the number of $\ol{u}$ elements. However,
arising expressions are too long. Therefore, we choose a more
intuitive approach.}}

According to Lemma 3,
$$
\begin{array}{l}
\MB{T}^{(+)}_0(x)\Bigl<\MB{B}_1(u),\MB{\Phi}\Bigr>=
f(x^{-1}u)\Bigl<\MB{B}_1(u),\wh{\MB{T}}^{(+)}_{0;1}(x;u)\MB{\Phi}\Bigr>+
g(xu^{-1})\Bigl<\MB{B}_1(x),\wh{\MB{T}}^{(+)}_{0;1}(u;u)\MB{\Phi}\Bigr>\\[6pt]
\MB{T}^{(-)}_0(x)\Bigl<\MB{B}_1(u),\MB{\Phi}\Bigr>=
f(xu^{-1})\Bigl<\MB{B}_1(u),\wh{\MB{T}}^{(-)}_{0;1}(x;u)\MB{\Phi}\Bigr>-
g(xu^{-1})\Bigl<\MB{B}_1(x),\wh{\MB{T}}^{(-)}_{0;1}(u;u)\MB{\Phi}\Bigr>.
\end{array}
$$
Hence, it follows
$$
\begin{array}{l}
\MB{T}^{(+)}_0(x)\Bigl<\MB{B}_{1,\ldots,M}(\vec{u}),\MB{\Phi}\Bigr>_{1,\ldots,M}=
\MB{T}^{(+)}_0(x)\Bigl<\MB{B}_1(u_1),\bigl<\MB{B}_{2,\ldots,M}(\vec{u}_1),\MB{\Phi}\bigr>_{2,\ldots,M}\Bigr>_1=\\[6pt]
\hskip10mm=
f(x^{-1}u_1)\Bigl<\MB{B}_1(u_1),\wh{\MB{T}}^{(+)}_{0;1}(x;u_1)
\bigl<\MB{B}_{2,\ldots,M}(\vec{u}_1),\MB{\Phi}\bigr>_{2,\ldots,M}\Bigr>_1+\\[4pt]
\hskip30mm+
g(xu_1^{-1})\Bigl<\MB{B}_1(x),\wh{\MB{T}}^{(+)}_{0;1}(u_1;u_1)
\bigl<\MB{B}_{2,\ldots,M}(\vec{u}_1),\MB{\Phi}\bigr>_{2,\ldots,M}\Bigr>_1\\[9pt]
\MB{T}^{(-)}_0(x)\Bigl<\MB{B}_{1,\ldots,M}(\vec{u}),\MB{\Phi}\Bigr>_{1,\ldots,M}=
\MB{T}^{(-)}_0(x)\Bigl<\MB{B}_1(u_1),\bigl<\MB{B}_{2,\ldots,M}(\vec{u}_1),\MB{\Phi}\bigr>_{2,\ldots,M}\Bigr>_1=\\[6pt]
\hskip10mm=
f(x^{-1}u_1)\Bigl<\MB{B}_1(u_1),\wh{\MB{T}}^{(-)}_{0;1}(x;u_1)
\bigl<\MB{B}_{2,\ldots,M}(\vec{u}_1),\MB{\Phi}\bigr>_{2,\ldots,M}\Bigr>_1-\\[4pt]
\hskip30mm-
g(xu_1^{-1})\Bigl<\MB{B}_1(x),\wh{\MB{T}}^{(-)}_{0;1}(u_1;u_1)
\bigl<\MB{B}_{2,\ldots,M}(\vec{u}_1),\MB{\Phi}\bigr>_{2,\ldots,M}\Bigr>_1
\end{array}
$$
With these relations it is easy to see that the following relations
$$
\begin{array}{l}
\MB{T}^{(+)}_0(x)
\Bigl<\MB{B}_{1,\ldots,M}(\vec{u}),\MB{\Phi}\Bigr>=
F(x^{-1},\ol{u})\Bigl<\MB{B}_{1,\ldots,M}(\vec{u}),
\wh{\MB{T}}^{(+)}_{0;1\ldots,M}(x;\vec{u})\MB{\Phi}\Bigr>+\\[6pt]
\hskip20mm+ g(xu_1^{-1})F(u_1;\ol{u}_1)
\Bigl<\MB{B}_{1;1\ldots,M}(x;\vec{u}_1),
\wh{\MB{T}}^{(+)}_{0;1\ldots,M}(u_1;\vec{u})\MB{\Phi}\Bigr>+\\[6pt]
\hskip40mm \mbox{$+$ terms without
$\MB{B}_{1;1\ldots,M}(x;\vec{u}_1)$}\\[9pt]
\MB{T}^{(-)}_0(x)\Bigl<\MB{B}_{1,\ldots,M}(\vec{u}),\MB{\Phi}\Bigr>=
F(x;\ol{u}^{-1})\Bigl<\MB{B}_{1,\ldots,M}(\vec{u}),
\wh{\MB{T}}^{(-)}_{0;1,\ldots,M}(x;\vec{u})\MB{\Phi}\Bigr>-\\[6pt]
\hskip20mm- g(xu_1^{-1})F(u_1;\ol{u}_1^{-1})\Bigr)
\Bigl<\MB{B}_{1;1,\ldots,M}(x;\vec{u}_1),
\wh{\MB{T}}^{(-)}_{0;1,\ldots,M}(u_1;\vec{u})\MB{\Phi}\Bigr>+\\[6pt]
\hskip40mm \mbox{$+$ terms without
$\MB{B}_{1;1\ldots,M}(x;\vec{u}_1)$}
\end{array}
$$
are true.

To get terms that contain $\MB{B}_{k;1\ldots,M}(x;\vec{u}_k)$ also
for $k>1$, we commute $\MB{B}_k(u_k)$ in the first place in
$\MB{B}_{1,\ldots,M}(\vec{u})$.

The commutation relations (\ref{RTT-1}) in the RTT--algebra of
$U_q\bigl(\MR{sp}(2n)\bigr)$ type
$$
\begin{array}{l}
T^i_{-k}(x)T^r_{-s}(y)+
\delta_{i,r}\Bigl(a(xy^{-1})-1\Bigr)T^i_{-k}(x)T^r_{-s}(y)+\\[6pt]
\hskip20mm+ b(xy^{-1})\theta_{r-i}T^r_{-k}(x)T^i_{-s}(y)-
b(x^{-1}y)\theta_{i-r}T^r_{-k}(x)T^i_{-s}(y)=\\[9pt]
\hskip10mm= T^r_{-s}(y)T^i_{-k}(x)+
\delta_{k,s}\Bigl(a(xy^{-1})-1\Bigr)T^r_{-s}(y)T^i_{-k}(x)+\\[6pt]
\hskip20mm+ b(xy^{-1})\theta_{s-k}T^r_{-k}(y)T^i_{-s}(x)-
b(x^{-1}y)\theta_{k-s}T^r_{-k}(y)T^i_{-s}(x)
\end{array}
$$
can be written in the matrix form as
$$
\MB{R}^{(+,+)}_{1,2}(xy^{-1})
\tilde{\MB{B}}_1(x)\tilde{\MB{B}}_2(y)=
\tilde{\MB{B}}_2(y)\tilde{\MB{B}}_1(x) \MB{R}^{(-,-)}_{1,2}(xy^{-1})
$$
or
$$
\tilde{\MB{B}}_1(x)\tilde{\MB{B}}_2(y)=
\bigl(\MB{R}^{(+,+)}_{1,2}(xy^{-1})\bigr)^{-1}
\tilde{\MB{B}}_2(y)\tilde{\MB{B}}_1(x) \MB{R}^{(-,-)}_{1,2}(xy^{-1})
$$
In our formalism, this is the relationship
$$
\begin{array}{l}
\Bigl<\MB{B}_1(x)\otimes\MB{B}_2(y),\MB{\Phi}\Bigr>=
\Bigl<\MB{B}_2(y)\otimes\MB{B}_1(x),
\bigl(\wh{\MB{R}}^{(+,+)}_{1^*,2^*}(xy^{-1})\bigr)^{-1}
\wh{\MB{R}}^{(-,-)}_{1,2}(xy^{-1}) \MB{\Phi}\Bigr>
\end{array}
$$
Gradually applying this relationship, we find that it holds
$$
\Bigl<\MB{B}_{1,\ldots,M}(\vec{u}),\MB{\Phi}\Bigr>=
\Bigl<\MB{B}_{k;1,\ldots,M}(u_k;\vec{u}_k),
\bigl(\wh{\MB{R}}^{(+,+)}_{1^*,\ldots,k^*}(\vec{u})\bigr)^{-1}
\wh{\MB{R}}^{(-,-)}_{1,\ldots,k}(\vec{u})\MB{\Phi}\Bigr>
$$
As for $k=1$, we get
$$
\begin{array}{l}
\MB{T}^{(+)}_0(x)\Bigl<\MB{B}_{1,\ldots,M}(\vec{u}),\MB{\Phi}\Bigr>=
\MB{T}^{(+)}_0(x)\Bigl<\MB{B}_{k;1,\ldots,M}(u_k;\vec{u}_k),
\bigl(\wh{\MB{R}}^{(+,+)}_{1^*,\ldots,k^*}(\vec{u})\bigr)^{-1}
\wh{\MB{R}}^{(-,-)}_{1,\ldots,k}(\vec{u})\MB{\Phi}\Bigr>=\\[6pt]
\hskip10mm=
F(x^{-1};\ol{u})\Bigl<\MB{B}_{k;1,\ldots,M}(u_k;\vec{u}_k),
\bigl(\wh{\MB{R}}^{(+,+)}_{0,k^*}(xu_k^{-1})\bigr)^{-1}
\wh{\MB{T}}^{(+)}_{0;1\ldots,\wh{k},\ldots,M}(x;\vec{u}_k)\\[4pt]
\hskip40mm \wh{\MB{R}}^{(+,-)}_{0,k}(xu_k^{-1})
\bigl(\wh{\MB{R}}^{(+,+)}_{1^*,\ldots,k^*}(\vec{u})\bigr)^{-1}
\wh{\MB{R}}^{(-,-)}_{1,\ldots,k}(\vec{u})\MB{\Phi}\Bigr>+\\[4pt]
\hskip20mm+
g(xu_k^{-1})F(u_k^{-1};\ol{u}_k)\Bigl<\MB{B}_{k;1,\ldots,M}(x;\vec{u}_k),
\bigl(\wh{\MB{R}}^{(+,+)}_{0,k^*}(1)\bigr)^{-1}
\wh{\MB{T}}^{(+)}_{0;1,\ldots,\wh{k},\ldots,M}(u_k;\vec{u}_k)\\[4pt]
\hskip40mm \wh{\MB{R}}^{(+,-)}_{0,k}(1)
\bigl(\wh{\MB{R}}^{(+,+)}_{1^*,\ldots,k^*}(\vec{u})\bigr)^{-1}
\wh{\MB{R}}^{(-,-)}_{1,\ldots,k}(\vec{u})\MB{\Phi}\Bigr>+\\[6pt]
\hskip40mm \mbox{$+$ terms without
$\MB{B}_{k;1,\ldots,M}(x;\vec{u}_k)$}\\[9pt]
\MB{T}^{(-)}_0(x)\Bigl<\MB{B}_{1,\ldots,M}(\vec{u}),\MB{\Phi}\Bigr>=
\MB{T}^{(-)}_0(x)\Bigl<\MB{B}_{k;1,\ldots,M}(u_k;\vec{u}_k),
\bigl(\wh{\MB{R}}^{(+,+)}_{1^*,\ldots,k^*}(\vec{u})\bigr)^{-1}
\wh{\MB{R}}^{(-,-)}_{1,\ldots,k}(\vec{u})\MB{\Phi}\Bigr>=\\[6pt]
\hskip10mm=
F(x^{-1};\ol{u})\Bigl<\MB{B}_{k;1,\ldots,M}(u_k;\vec{u}_k),
\bigl(\wh{\MB{R}}^{(-,+)}_{0,k^*}(xu_k^{-1})\bigr)^{-1}
\wh{\MB{T}}^{(-)}_{0;1\ldots,\wh{k},\ldots,M}(x;\vec{u}_k)\\[4pt]
\hskip40mm \wh{\MB{R}}^{(-,-)}_{0,k}(xu_k^{-1})
\bigl(\wh{\MB{R}}^{(+,+)}_{1^*,\ldots,k^*}(\vec{u})\bigr)^{-1}
\wh{\MB{R}}^{(-,-)}_{1,\ldots,k}(\vec{u})\MB{\Phi}\Bigr>-\\[4pt]
\hskip20mm-
g(xu_k^{-1})F(u_k^{-1};\ol{u}_k)\Bigl<\MB{B}_{k;1,\ldots,M}(x;\vec{u}_k),
\bigl(\wh{\MB{R}}^{(-,+)}_{0,k^*}(1)\bigr)^{-1}
\wh{\MB{T}}^{(-)}_{0;1,\ldots,\wh{k},\ldots,M}(u_k;\vec{u}_k)\\[4pt]
\hskip40mm \wh{\MB{R}}^{(-,-)}_{0,k}(1)
\bigl(\wh{\MB{R}}^{(+,+)}_{1^*,\ldots,k^*}(\vec{u})\bigr)^{-1}
\wh{\MB{R}}^{(-,-)}_{1,\ldots,k}(\vec{u})\MB{\Phi}\Bigr>+\\[6pt]
\hskip40mm \mbox{$+$ terms without
$\MB{B}_{k;1,\ldots,M}(x;\vec{u}_k)$}
\end{array}
$$
Using the definitions of the given operators and the Yang--Baxter
equations,
$$
\begin{array}{l}
\wh{\MB{R}}^{(\pm,+)}_{1,3^*}(xy)\wh{\MB{R}}^{(\pm,+)}_{1,2^*}(x)\wh{\MB{R}}^{(+,+)}_{2^*,3^*}(y)=
\wh{\MB{R}}^{(+,+)}_{2^*,3^*}(y)\wh{\MB{R}}^{(\pm,+)}_{1,2^*}(x)\wh{\MB{R}}^{(\pm,+)}_{1,3^*}(xy)\\[4pt]
\wh{\MB{R}}^{(\pm,-)}_{1,2}(x)\wh{\MB{R}}^{(\pm,-)}_{1,3}(xy)\wh{\MB{R}}^{(-,-)}_{2,3}(y)=
\wh{\MB{R}}^{(-,-)}_{2,3}(y)\wh{\MB{R}}^{(\pm,-)}_{1,3}(xy)\wh{\MB{R}}^{(\pm,-)}_{1,2}(x)
\end{array}
$$
it is relatively easy to show that the following relations
$$
\begin{array}{l}
\MB{T}^{(+)}_0(x)\Bigl<\MB{B}_{1,\ldots,M}(\vec{u}),\MB{\Phi}\Bigr>=
F(x^{-1};\ol{u})\Bigl<\MB{B}_{1,\ldots,M}(\vec{u}),
\wh{\MB{T}}^{(+)}_{0;1,\ldots,M}(x;\vec{u})\MB{\Phi}\Bigr>+\\[6pt]
\hskip10mm+
g(xu_k^{-1})F(u_k^{-1};\ol{u}_k)\Bigl<\MB{B}_{k;1,\ldots,M}(x;\vec{u}_k),
\bigl(\wh{\MB{R}}^{(+,+)}_{1^*,\ldots,k^*}(\vec{u})\bigr)^{-1}
\wh{\MB{R}}^{(-,-)}_{1,\ldots,k}(\vec{u})
\wh{\MB{T}}^{(+)}_{0;1,\ldots,M}(u_k;\vec{u})\MB{\Phi}\Bigr>+\\[6pt]
\hskip40mm \mbox{$+$ terms without
$\MB{B}_{k;1,\ldots,M}(x;\vec{u}_k)$}\\[9pt]
\MB{T}^{(-)}_0(x)\Bigl<\MB{B}_{1,\ldots,M}(\vec{u}),\MB{\Phi}\Bigr>=
F(x;\ol{u}^{-1})\Bigl<\MB{B}_{1,\ldots,M}(\vec{u}),
\wh{\MB{T}}^{(-)}_{0;1,\ldots,M}(x;\vec{u})\MB{\Phi}\Bigr>-\\[6pt]
\hskip10mm-
g(xu_k^{-1})F(u_k;\ol{u}_k^{-1}\Bigl<\MB{B}_{k;1,\ldots,M}(x;\vec{u}_k),
\bigl(\wh{\MB{R}}^{(+,+)}_{1^*,\ldots,k^*}(\vec{u})\bigr)^{-1}
\wh{\MB{R}}^{(-,-)}_{1,\ldots,k}(\vec{u})
\wh{\MB{T}}^{(-)}_{0;1,\ldots,M}(u_k;\vec{u})\MB{\Phi}\Bigr>+\\[6pt]
\hskip40mm \mbox{$+$ terms without
$\MB{B}_{k;1,\ldots,M}(x;\vec{u}_k)$}
\end{array}
$$
are true. This proves our claim.

\subsection*{A6\quad Proof of Theorem 1}

Following Proposition 2, we have
$$
\begin{array}{l}
\Tr\bigl(\MB{T}^{(+)}_0(x)\bigr)\Bigl<\MB{B}_{1,\ldots,M}(\vec{u}),\MB{\Phi}\Bigr>=
F(x^{-1};\ol{u})\Bigl<\MB{B}_{1,\ldots,M}(\vec{u}),
\Tr\bigl(\wh{\MB{T}}^{(+)}_{0;1,\ldots,M}(x;\vec{u})\bigr)_0\MB{\Phi}\Bigr>+\\[6pt]
\hskip20mm+
{\tsum_{u_k\in\ol{u}}}g(xu_k^{-1})F(u_k^{-1};\ol{u}_k)
\Bigl<\MB{B}_{k;1,\ldots,M}(x;\vec{u}_k),
\bigl(\wh{\MB{R}}^{(+,+)}_{1^*,\ldots,k^*}(\vec{u})\bigr)^{-1}
\wh{\MB{R}}^{(-,-)}_{1,\ldots,k}(\vec{u})\\[6pt]
\hskip60mm
\Tr\bigl(\wh{\MB{T}}^{(+)}_{0;1,\ldots,M}(u_k;\vec{u})\bigr)_0\MB{\Phi}\Bigr>\\[12pt]
\Tr\bigl(\MB{T}^{(-)}_0(x)\bigr)\Bigl<\MB{B}_{1,\ldots,M}(\vec{u}),\MB{\Phi}\Bigr>=
F(x;\ol{u}^{-1})\Bigl<\MB{B}_{1,\ldots,M}(\vec{u}),
\Tr\bigl(\wh{\MB{T}}^{(-)}_{0;1,\ldots,M}(x;\vec{u})\bigr)_0\MB{\Phi}\Bigr>-\\[6pt]
\hskip20mm-
{\tsum_{u_k\in\ol{u}}}g(xu_k^{-1})
F(u_k;\ol{u}_k^{-1}\Bigl<\MB{B}_{k;1,\ldots,M}(x;\vec{u}_k),
\bigl(\wh{\MB{R}}^{(+,+)}_{1^*,\ldots,k^*}(\vec{u})\bigr)^{-1}
\wh{\MB{R}}^{(-,-)}_{1,\ldots,k}(\vec{u})\\[6pt]
\hskip60mm
\Tr\bigl(\wh{\MB{T}}^{(-)}_{0;1,\ldots,M}(u_k;\vec{u})\bigr)_0\MB{\Phi}\Bigr>
\end{array}
$$
Therefore, for the eigenvector $\MB{\Phi}$ we obtain
$$
\begin{array}{l}
\Tr\bigl(\MB{T}^{(+)}_0(x)\bigr)\Bigl<\MB{B}_{1,\ldots,M}(\vec{u}),\MB{\Phi}\Bigr>=
\wh{E}^{(+)}_{1,\ldots,M}(x;\vec{u})F(x^{-1};\ol{u})
\Bigl<\MB{B}_{1,\ldots,M}(\vec{u}),\MB{\Phi}\Bigr>+\\[6pt]
\hskip10mm+
{\tsum_{u_k\in\ol{u}}}
\wh{E}^{(+)}_{1,\ldots,M}(u_k;\vec{u})
F(u_k^{-1};\ol{u}_k)g(xu_k^{-1})
\Bigl<\MB{B}_{k;1,\ldots,M}(x;\vec{u}_k),
\bigl(\wh{\MB{R}}^{(+,+)}_{1^*,\ldots,k^*}(\vec{u})\bigr)^{-1}
\wh{\MB{R}}^{(-,-)}_{1,\ldots,k}(\vec{u})\MB{\Phi}\Bigr>\\[12pt]
\Tr\bigl(\MB{T}^{(-)}_0(x)\bigr)\Bigl<\MB{B}_{1,\ldots,M}(\vec{u}),\MB{\Phi}\Bigr>=
\wh{E}^{(-)}_{1,\ldots,M}(x;\vec{u})F(x;\ol{u}^{-1})
\Bigl<\MB{B}_{1,\ldots,M}(\vec{u}),\MB{\Phi}\Bigr>-\\[6pt]
\hskip10mm-
{\tsum_{u_k\in\ol{u}}}\wh{E}^{(-)}_{1,\ldots,M}(u_k;\vec{u})
F(u_k;\ol{u}_k^{-1}g(xu_k^{-1})\Bigl<\MB{B}_{k;1,\ldots,M}(x;\vec{u}_k),
\bigl(\wh{\MB{R}}^{(+,+)}_{1^*,\ldots,k^*}(\vec{u})\bigr)^{-1}
\wh{\MB{R}}^{(-,-)}_{1,\ldots,k}(\vec{u})\MB{\Phi}\Bigr>
\end{array}
$$
and thus the relation
$$
\begin{array}{l}
H(x)\Bigl<\MB{B}_{1,\ldots,M}(\vec{u}),\MB{\Phi}\Bigr>=
\Tr\bigl(\MB{T}^{(+)}_0(x)\bigr)\Bigl<\MB{B}_{1,\ldots,M}(\vec{u}),\MB{\Phi}\Bigr>+
\Tr\bigl(\MB{T}^{(-)}_0(x)\bigr)\Bigl<\MB{B}_{1,\ldots,M}(\vec{u}),\MB{\Phi}\Bigr>=\\[6pt]
\hskip10mm=
\Bigl(\wh{E}^{(+)}_{1,\ldots,M}(x;\vec{u})F(x^{-1};\ol{u})+
\wh{E}^{(-)}_{1,\ldots,M}(x;\vec{u})F(x;\ol{u}^{-1})\Bigr)
\Bigl<\MB{B}_{1,\ldots,M}(\vec{u}),\MB{\Phi}\Bigr>+\\[9pt]
\hskip20mm+
{\tsum_{u_k\in\ol{u}}}
\Bigl(\wh{E}^{(+)}_{1,\ldots,M}(u_k;\vec{u})F(u_k^{-1};\ol{u}_k)-
\wh{E}^{(-)}_{1,\ldots,M}(u_k;\vec{u})F(u_k;\ol{u}_k^{-1}\Bigr)g(xu_k^{-1})\\[6pt]
\hskip40mm
\Bigl<\MB{B}_{k;1,\ldots,M}(x;\vec{u}_k),
\bigl(\wh{\MB{R}}^{(+,+)}_{1^*,\ldots,k^*}(\vec{u})\bigr)^{-1}
\wh{\MB{R}}^{(-,-)}_{1,\ldots,k}(\vec{u})\MB{\Phi}\Bigr>
\end{array}
$$
is valid.

So if the condition (\ref{BP-1}) is fulfilled,
$\Bigl<\MB{B}_{1,\ldots,M}(\vec{u}),\MB{\Phi}\Bigr>$ is the
eigenvector of the operator $H(x)$.

\subsection*{A7\quad Proof of Theorem 2}

The left-hand side of the proven equality can be written as
$$
\begin{array}{l}
\MB{R}^{(\epsilon,\epsilon')}_{0,0'}(xy^{-1})\wh{\MB{T}}^{(\epsilon)}_{0;1,\ldots,M}(x;\vec{u})
\wh{\MB{T}}^{(\epsilon')}_{0';1,\ldots,M}(y;\vec{u})=\\[9pt]
\hskip10mm= \MB{R}^{(\epsilon,\epsilon')}_{0,0'}(xy^{-1})
\Bigl(\bigl(\wh{\MB{R}}^{(\epsilon,+)}_{0,1^*}(xu_1^{-1})\bigr)^{-1}
\bigl(\wh{\MB{R}}^{(\epsilon',+)}_{0',1^*}(yu_1^{-1})\bigr)^{-1}\Bigr)\ldots\\[4pt]
\hskip20mm
\Bigl(\bigl(\wh{\MB{R}}^{(\epsilon,+)}_{0,M^*}(xu_M^{-1})\bigr)^{-1}
\bigl(\wh{\MB{R}}^{(\epsilon',+)}_{0',M^*}(yu_M^{-1})\bigr)^{-1}\Bigr)
\Bigl(\MB{T}^{(\epsilon)}_0(x)\MB{T}^{(\epsilon')}_{0'}(y)\Bigr)\\[4pt]
\hskip40mm \Bigl(\wh{\MB{R}}^{(\epsilon,-)}_{0,M}(xu_M^{-1})
\wh{\MB{R}}^{(\epsilon',-)}_{0',M}(yu_M^{-1})\Bigr)\ldots
\Bigl(\wh{\MB{R}}^{(\epsilon,-)}_{0,1}(xu_1^{-1})
\wh{\MB{R}}^{(\epsilon',-)}_{0',1}(yu_1^{-1})\Bigr)
\end{array}
$$
The statement of the Theorem then results from the relations
$$
\begin{array}{l}
\MB{R}^{(\epsilon,\epsilon')}_{0,0'}(x)
\wh{\MB{R}}^{(\epsilon,+)}_{0',1^*}(y)
\wh{\MB{R}}^{(\epsilon',+)}_{0,1^*}(xy)=
\wh{\MB{R}}^{(\epsilon',+)}_{0,1^*}(xy)
\wh{\MB{R}}^{(\epsilon,+)}_{0',1^*}(y)
\MB{R}^{(\epsilon,\epsilon')}_{0,0'}(x)\\[4pt]
\MB{R}^{(\epsilon,\epsilon')}_{0,0'}(x)
\wh{\MB{R}}^{(\epsilon,-)}_{0,1}(xy)
\wh{\MB{R}}^{(\epsilon',-)}_{0',1}(y)=
\wh{\MB{R}}^{(\epsilon',-)}_{0',1}(y)
\wh{\MB{R}}^{(\epsilon,-)}_{0,1}(xy)
\MB{R}^{(\epsilon,\epsilon')}_{0,0'}(x)
\end{array}
$$
which are a direct result of the Yang--Baxter equations
(\ref{YB-epsilon}) for the RTT--algebra $\tilde{\MC{A}}_n$ and
RTT--equations (\ref{RTT-epsilon}).

\subsection*{A8\quad Proof of Theorem 3}

First, we calculate
$\wh{T}^i_k(x;\vec{u})\wh{\MB{\Omega}}$ for $i,\,k>0$. If we write
$$
\begin{array}{l}
\bigl(\wh{\MB{R}}^{(+,+)}_{0,1^*}(x)\Bigr)^{-1}=
{\tsum_{r,s,a,b=1}^n}\wh{R}^{r,a}_{s,b}(x)\MB{E}^s_r\otimes\MB{F}^b_a\otimes\MB{I}_-\,,\\[9pt]
\wh{\MB{R}}^{(+,-)}_{0,1}(x)=
{\tsum_{p,q,c,d=1}^n}\wh{R}^{p,-c}_{q,-d}(x)\MB{E}^q_p\otimes\MB{I}^*_+\otimes\MB{E}^{-d}_{-c}\,,
\end{array}
$$
where
$$
\begin{array}{l}
\wh{R}^{r,a}_{s,b}(x)= \dfrac1{f(x^{-1})}\Bigl(
\delta^r_s\delta^a_b\bigl(1+\delta_{r,a}(f(x^{-1})-1)\bigr)+
\delta^r_b\delta^a_s\bigl(g(x^{-1})\theta_{r-s}-g(x)\theta_{s-r}\bigr)\Bigr)\\[9pt]
\wh{R}^{p,-c}_{q,-d}(x)= \delta^p_q\delta^c_d
\Bigl(1+\delta_{p,c}\bigl(f(x^{-1}q)-1\bigr)\Bigr)+
\delta^{p,c}\delta_{q,d}q^{p-q}
\Bigl(g(x^{-1}q)\theta_{p-q}-g(xq^{-1})\theta_{q-p}\Bigr),
\end{array}
$$
we get
$$
\begin{array}{l}
\wh{\MB{T}}^{(+)}_{0;1,\ldots,M}(x;\vec{u})=
\wh{R}^{r_1,a_1}_{s_1,b_1}(xu_1^{-1})
\wh{R}^{r_2,a_2}_{s_2,b_2}(xu_2^{-1})\ldots
\wh{R}^{r_M,a_M}_{s_M,b_M}(xu_M^{-1})\\[6pt]
\hskip20mm \wh{R}^{p_M,-c_M}_{q_M,-d_M}(xu_M^{-1})\ldots
\wh{R}^{p_2,-c_2}_{q_2,-d_2}(xu_2^{-1})
\wh{R}^{p_1,-c_1}_{q_1,-d_1}(xu_1^{-1})\\[6pt]
\hskip20mm
\MB{E}^{s_1}_{r_1}\MB{E}^{s_2}_{r_2}\ldots\MB{E}^{s_M}_{r_M}
\MB{E}^u_v\MB{E}^{q_M}_{p_M}\ldots\MB{E}^{q_2}_{p_2}\MB{E}^{q_1}_{p_1}\otimes\\[6pt]
\hskip30mm
\MB{F}^{b_1}_{a_1}\otimes\MB{F}^{b_2}_{a_2}\otimes\ldots\otimes\MB{F}^{b_M}_{a_M}\otimes
\MB{E}^{-d_1}_{-c_1}\otimes\MB{E}^{-d_2}_{-c_2}\otimes\ldots\otimes\MB{E}^{-d_M}_{-c_M}\otimes
T^v_u(x)
\end{array}
$$
This leads to the relationship
$$
\begin{array}{l}
\wh{T}^i_k(x;\vec{u})= \wh{R}^{i,a_1}_{s_1,b_1}(xu_1^{-1})
\wh{R}^{s_1,a_2}_{s_2,b_2}(xu_2^{-1})\ldots
\wh{R}^{s_{M-1},a_M}_{s_M,b_M}(xu_M^{-1})\\[6pt]
\hskip20mm \wh{R}^{p_M,-c_M}_{p_{M-1},-d_M}(xu_M^{-1})\ldots
\wh{R}^{p_2,-c_2}_{p_1,-d_2}(xu_2^{-1})
\wh{R}^{p_1,-c_1}_{k,-d_1}(xu_1^{-1})\\[6pt]
\hskip20mm
\MB{F}^{b_1}_{a_1}\otimes\MB{F}^{b_2}_{a_2}\otimes\ldots\otimes\MB{F}^{b_M}_{a_M}\otimes
\MB{E}^{-d_1}_{-c_1}\otimes\MB{E}^{-d_2}_{-c_2}\otimes\ldots\otimes\MB{E}^{-d_M}_{-c_M}\otimes
T^{s_M}_{p_M}(x)
\end{array}
$$
and thus
$$
\begin{array}{l}
\wh{T}^i_k(x;\vec{u})\wh{\MB{\Omega}}=
\wh{R}^{i,1}_{s_1,b_1}(xu_1^{-1})
\wh{R}^{s_1,1}_{s_2,b_2}(xu_2^{-1})\ldots
\wh{R}^{s_{M-1},1}_{s_M,b_M}(xu_M^{-1})\\[6pt]
\hskip20mm \wh{R}^{p_M,-c_M}_{p_{M-1},-1}(xu_M^{-1})\ldots
\wh{R}^{p_2,-c_2}_{p_1,-1}(xu_2^{-1})
\wh{R}^{p_1,-c_1}_{k,-1}(xu_1^{-1})\\[6pt]
\hskip20mm
\MB{f}^{b_1}\otimes\MB{f}^{b_2}\otimes\ldots\otimes\MB{f}^{b_M}\otimes
\MB{e}_{-c_1}\otimes\MB{e}_{-c_2}\otimes\ldots\otimes\MB{e}_{-c_M}\otimes
T^{s_M}_{p_M}(x)\omega .
\end{array}
$$
For $k>1$  $\wh{R}^{p,-c}_{k,-1}(x)=\delta^p_k\delta^c_1$, and so
$$
\begin{array}{l}
\wh{T}^i_k(x;\vec{u})\wh{\MB{\Omega}}=
\wh{R}^{i,1}_{s_1,b_1}(xu_1^{-1})
\wh{R}^{s_1,1}_{s_2,b_2}(xu_2^{-1})\ldots
\wh{R}^{s_{M-1},1}_{s_M,b_M}(xu_M^{-1})\\[6pt]
\hskip20mm
\MB{f}^{b_1}\otimes\MB{f}^{b_2}\otimes\ldots\otimes\MB{f}^{b_M}\otimes
\MB{e}_{-1}\otimes\MB{e}_{-1}\otimes\ldots\otimes\MB{e}_{-1}\otimes
T^{s_M}_k(x)\omega .
\end{array}
$$
Since $T^{s_M}_k(x)\omega=0$ for $s_M<k$, we have $k=2,\,\ldots,s_M$, and therefore
$$
\wh{R}^{s_{M-1},1}_{s_M,b_M}(xu_M^{-1})=
\dfrac1{f(x^{-1}u_M)}\,\delta^{s_{M-1}}_{s_M}\delta^1_{b_M}=
f(xu_M^{-1}q^{-1})\delta^{s_{M-1}}_{s_M}\delta^1_{b_M}
$$
Using this relationship, we gradually get
$$
\wh{T}^i_k(x;\vec{u})\wh{\MB{\Omega}}= F(xq^{-1};\ol{u}^{-1})
\MB{f}^1\otimes\MB{f}^1\otimes\ldots\otimes\MB{f}^1\otimes
\MB{e}_{-1}\otimes\MB{e}_{-1}\otimes\ldots\otimes\MB{e}_{-1}\otimes
T^i_k(x)\omega
$$
Since $\omega$ is a vacuum vector, we get  for $k\geq2$
$$
\wh{T}^i_k(x;\vec{u})\wh{\MB{\Omega}}=0\quad\MR{for}\quad i<k\,,
\hskip15mm \wh{T}^k_k(x;\vec{u})\wh{\MB{\Omega}}=
\lambda_k(x)F(xq^{-1};\ol{u}^{-1})\wh{\MB{\Omega}}
$$

\bigskip

For $i=k=1$ we have
$$
\begin{array}{l}
\wh{T}^1_1(x;\vec{u})\wh{\MB{\Omega}}=
\wh{R}^{1,1}_{s_1,b_1}(xu_1^{-1})
\wh{R}^{s_1,1}_{s_2,b_2}(xu_2^{-1})\ldots
\wh{R}^{s_{M-1},1}_{s_M,b_M}(xu_M^{-1})\\[6pt]
\hskip20mm \wh{R}^{p_M,-c_M}_{p_{M-1},-1}(xu_M^{-1})\ldots
\wh{R}^{p_2,-c_2}_{p_1,-1}(xu_2^{-1})
\wh{R}^{p_1,-c_1}_{1,-1}(xu_1^{-1})\\[6pt]
\hskip20mm
\MB{f}^{b_1}\otimes\MB{f}^{b_2}\otimes\ldots\otimes\MB{f}^{b_M}\otimes
\MB{e}_{-c_1}\otimes\MB{e}_{-c_2}\otimes\ldots\otimes\MB{e}_{-c_M}\otimes
T^{s_M}_{p_M}(x)\omega
\end{array}.
$$
Since $\wh{R}^{1,1}_{s,b}(xu^{-1})=\delta^1_s\delta^1_b$, we obtain
$$
\begin{array}{l}
\wh{T}^1_1(x;\vec{u})\wh{\MB{\Omega}}=
\wh{R}^{p_M,-c_M}_{p_{M-1},-1}(xu_M^{-1})\ldots
\wh{R}^{p_2,-c_2}_{p_1,-1}(xu_2^{-1})
\wh{R}^{p_1,-c_1}_{1,-1}(xu_1^{-1})\\[6pt]
\hskip20mm \MB{f}^1\otimes\MB{f}^1\otimes\ldots\otimes\MB{f}^\otimes
\MB{e}_{-c_1}\otimes\MB{e}_{-c_2}\otimes\ldots\otimes\MB{e}_{-c_M}\otimes
T^1_{p_M}(x)\omega\,.
\end{array}
$$
It is assumed that
$T^1_{p_M}(x)\omega=\lambda_1(x)\delta^1_{p_M}\omega$, and so
$$
\begin{array}{l}
\wh{T}^1_1(x;\vec{u})\wh{\MB{\Omega}}=
\lambda_1(x)\wh{R}^{1,-c_M}_{p_{M-1},-1}(xu_M^{-1})\ldots
\wh{R}^{p_2,-c_2}_{p_1,-1}(xu_2^{-1})
\wh{R}^{p_1,-c_1}_{1,-1}(xu_1^{-1})\\[6pt]
\hskip20mm \MB{f}^1\otimes\MB{f}^1\otimes\ldots\otimes\MB{f}^\otimes
\MB{e}_{-c_1}\otimes\MB{e}_{-c_2}\otimes\ldots\otimes\MB{e}_{-c_M}\otimes\omega .
\end{array}
$$
Since $\wh{R}^{1,-c}_{p,-1}(xu^{-1})=
\delta^1_p\delta^c_1f(x^{-1}uq)$, we have
$$
\wh{T}^1_1(x;\vec{u})\wh{\MB{\Omega}}=
\lambda_1(x)F(x^{-1}q;\ol{u})\wh{\MB{\Omega}}.
$$

\bigskip

In a similar way for the negative part we write
$$
\begin{array}{l}
\bigl(\wh{\MB{R}}^{(-,+)}_{0,1^*}(x)\bigr)^{-1}=
{\tsum_{r,s,a,b=1}^n}\wh{R}^{-r,a}_{-s,b}(x)\MB{E}^{-s}_{-r}\otimes\MB{F}^b_a\otimes\MB{I}_-\,,\\[9pt]
\wh{\MB{R}}^{(-,-)}_{0,1}(x)=
{\tsum_{p,q,c,d=1}^n}\wh{R}^{-p,-c}_{-q,-d}(x)\MB{E}^{-q}_{-p}\otimes\MB{I}^*_+\otimes\MB{E}^{-d}_{-c}
\end{array}
$$
where
$$
\begin{array}{l}
\wh{R}^{-r,a}_{-s,b}(x)=
\delta^r_s\delta^a_b\Bigl(1+\delta_{r,a}\bigl(f(xq)-1\bigr)\Bigr)+
\delta^{r,a}\delta_{s,b}q^{r-s}\Bigl(g(xq)\theta_{r-s}-g(x^{-1}q^{-1})\theta_{s-r}\Bigr)\\[6pt]
\wh{R}^{-p,-c}_{-q,-d}(x)=
\dfrac1{f(x)}\Bigl(\delta^p_q\delta^c_d\bigl(1+\delta_{p,c}(f(x)-1)\bigr)+
\delta^p_d\delta^c_q\bigl(g(x)\theta_{p-q}-g(x^{-1})\theta_{q-p}\bigr)\Bigr) .
\end{array}
$$
Then we get
$$
\begin{array}{l}
\wh{\MB{T}}^{(-)}_{0;1,\ldots,M}(x;\vec{u})=
\wh{R}^{-r_1,a_1}_{-s_1,b_1}(xu_1^{-1})
\wh{R}^{-s_1,a_2}_{-s_2,b_2}(xu_2^{-1})\ldots
\wh{R}^{-s_{M-1},a_M}_{-s_M,b_M}(xu_M^{-1})\\[6pt]
\hskip20mm
\wh{R}^{-p_M,-c_M}_{-q_M,-d_M}(xu_M^{-1})\wh{R}^{-q_M,-c_{M-1}}_{-q_{M-1},-d_{M-1}}(xu_{M-1}^{-1})\ldots
\wh{R}^{-q_2,-c_1}_{-q_1,-d_1}(xu_1^{-1})\\[6pt]
\hskip30mm
\MB{E}^{-q_1}_{-r_1}\otimes\MB{F}^{b_1}_{a_1}\otimes\MB{F}^{b_2}_{a_2}\otimes\ldots\otimes\MB{F}^{b_M}_{a_M}\otimes
\MB{E}^{-d_1}_{-c_1}\otimes\MB{E}^{-d_2}_{-c_2}\otimes\ldots\otimes\MB{E}^{-d_M}_{-c_M}\otimes
T^{-s_M}_{-p_M}(x)
\end{array}
$$
which means that
$$
\begin{array}{l}
\wh{T}^{-i}_{-k}(x;\vec{u})= \wh{R}^{-i,a_1}_{-s_1,b_1}(xu_1^{-1})
\wh{R}^{-s_1,a_2}_{-s_2,b_2}(xu_2^{-1})\ldots
\wh{R}^{-s_{M-1},a_M}_{-s_M,b_M}(xu_M^{-1})\\[6pt]
\hskip20mm
\wh{R}^{-p_M,-c_M}_{-q_M,-d_M}(xu_M^{-1})\wh{R}^{-q_M,-c_{M-1}}_{-q_{M-1},-d_{M-1}}(xu_{M-1}^{-1})\ldots
\wh{R}^{-q_2,-c_1}_{-q_1,-k}(xu_1^{-1})\\[6pt]
\hskip30mm
\MB{F}^{b_1}_{a_1}\otimes\MB{F}^{b_2}_{a_2}\otimes\ldots\otimes\MB{F}^{b_M}_{a_M}\otimes
\MB{E}^{-d_1}_{-c_1}\otimes\MB{E}^{-d_2}_{-c_2}\otimes\ldots\otimes\MB{E}^{-d_M}_{-c_M}\otimes
T^{-s_M}_{-p_M}(x) .
\end{array}
$$
Therefore we obtain
$$
\begin{array}{l}
\wh{T}^{-i}_{-k}(x;\vec{u})\wh{\MB{\Omega}}=
\wh{R}^{-i,1}_{-s_1,b_1}(xu_1^{-1})
\wh{R}^{-s_1,1}_{-s_2,b_2}(xu_2^{-1})\ldots
\wh{R}^{-s_{M-1},1}_{-s_M,b_M}(xu_M^{-1})\\[6pt]
\hskip20mm
\wh{R}^{-p_M,-c_M}_{-q_M,-1}(xu_M^{-1})\wh{R}^{-q_M,-c_{M-1}}_{-q_{M-1},-1}(xu_{M-1}^{-1})\ldots
\wh{R}^{-q_2,-c_1}_{-k,-1}(xu_1^{-1})\\[6pt]
\hskip30mm
\MB{f}^{b_1}\otimes\MB{f}^{b_2}\otimes\ldots\otimes\MB{f}^{b_M}\otimes
\MB{e}_{-c_1}\otimes\MB{e}_{-c_2}\otimes\ldots\otimes\MB{e}_{-c_M}\otimes
T^{-s_M}_{-p_M}(x)\omega
\end{array}
$$
For $i\geq2$ we have
$\wh{R}^{-i,1}_{-s,b}(xu^{-1})=\delta^i_s\delta^1_b$, and so
$$
\begin{array}{l}
\wh{T}^{-i}_{-k}(x;\vec{u})\wh{\MB{\Omega}}=
\wh{R}^{-p_M,-c_M}_{-q_M,-1}(xu_M^{-1})\wh{R}^{-q_M,-c_{M-1}}_{-q_{M-1},-1}(xu_{M-1}^{-1})\ldots
\wh{R}^{-q_2,-c_1}_{-k,-1}(xu_1^{-1})\\[6pt]
\hskip30mm
\MB{f}^1\otimes\MB{f}^1\otimes\ldots\otimes\MB{f}^1\otimes
\MB{e}_{-c_1}\otimes\MB{e}_{-c_2}\otimes\ldots\otimes\MB{e}_{-c_M}\otimes
T^{-i}_{-p_M}(x)\omega .
\end{array}
$$
Since $T^{-i}_{-p_M}(x)\omega=0$ for $i>p_M$, there must be $i=2,\,\ldots,\,p_M$.
But then we have
$$
\wh{R}^{-p_M,-c}_{-q,-1}(xu_M^{-1})=
\dfrac1{f(xu_M^{-1})}\delta^{p_M}_q\delta^c_1=
f(x^{-1}u_Mq^{-1})\delta^{p_M}_q\delta^c_1
$$
and hence
$$
\wh{T}^{-i}_{-k}(x;\vec{u})\wh{\MB{\Omega}}= F(x^{-1}q^{-1};\ol{u})
\MB{f}^1\otimes\MB{f}^1\otimes\ldots\otimes\MB{f}^1\otimes
\MB{e}_{-1}\otimes\MB{e}_{-1}\otimes\ldots\otimes\MB{e}_{-1}\otimes
T^{-i}_{-k}(x)\omega\,.
$$
Since $\omega$ is the vacuum vector, we get for $i\geq2$
$$
\wh{T}^{-i}_{-k}(x;\vec{u})\wh{\MB{\Omega}}=0\quad\MR{for}\quad
i>k\,,\hskip15mm \wh{T}^{-i}_{-i}(x;\vec{u})\wh{\MB{\Omega}}=
\lambda_{-i}(x)F(x^{-1}q^{-1};\ol{u})\wh{\MB{\Omega}}
$$
For $i=k=1$ we get
$$
\begin{array}{l}
\wh{T}^{-1}_{-1}(x;\vec{u})\wh{\MB{\Omega}}=
\wh{R}^{-1,1}_{-s_1,b_1}(xu_1^{-1})
\wh{R}^{-s_1,1}_{-s_2,b_2}(xu_2^{-1})\ldots
\wh{R}^{-s_{M-1},1}_{-s_M,b_M}(xu_M^{-1})\\[6pt]
\hskip20mm
\wh{R}^{-p_M,-c_M}_{-q_M,-1}(xu_M^{-1})\wh{R}^{-q_M,-c_{M-1}}_{-q_{M-1},-1}(xu_{M-1}^{-1})\ldots
\wh{R}^{-q_2,-c_1}_{-1,-1}(xu_1^{-1})\\[6pt]
\hskip30mm
\MB{f}^{b_1}\otimes\MB{f}^{b_2}\otimes\ldots\otimes\MB{f}^{b_M}\otimes
\MB{e}_{-c_1}\otimes\MB{e}_{-c_2}\otimes\ldots\otimes\MB{e}_{-c_M}\otimes
T^{-s_M}_{-p_M}(x)\omega .
\end{array}
$$
Since
$\wh{R}^{-1,1}_{-s,b}(xu^{-1})=f(xu^{-1}q)\delta^1_s\delta^1_b$,
$$
\begin{array}{l}
\wh{T}^{-1}_{-1}(x;\vec{u})\wh{\MB{\Omega}}= F(xq;\ol{u}^{-1})
\wh{R}^{-p_M,-c_M}_{-q_M,-1}(xu_M^{-1})\wh{R}^{-q_M,-c_{M-1}}_{-q_{M-1},-1}(xu_{M-1}^{-1})\ldots
\wh{R}^{-q_2,-c_1}_{-1,-1}(xu_1^{-1})\\[6pt]
\hskip30mm
\MB{f}^1\otimes\MB{f}^1\otimes\ldots\otimes\MB{f}^1\otimes
\MB{e}_{-c_1}\otimes\MB{e}_{-c_2}\otimes\ldots\otimes\MB{e}_{-c_M}\otimes
T^{-1}_{-p_M}(x)\omega .
\end{array}
$$
By definition we have
$\wh{R}^{-q,-c}_{-1,-1}(xu^{-1})=\delta^q_1\delta^c_1$, and so the following relations holds:
$$
\wh{T}^{-1}_{-1}(x;\vec{u})\wh{\MB{\Omega}}= F(xq;\ol{u}^{-1})
\MB{f}^1\otimes\MB{f}^1\otimes\ldots\otimes\MB{f}^1\otimes
\MB{e}_{-1}\otimes\MB{e}_{-1}\otimes\ldots\otimes\MB{e}_{-1}\otimes
T^{-1}_{-1}(x)\omega .
$$
And since $\omega$ is a vacuum vector, we hence get
$$
\wh{T}^{-1}_{-1}(x;\vec{u})\wh{\MB{\Omega}}=
\lambda_{-1}(x)F(xq;\ol{u}^{-1})\wh{\MB{\Omega}} .
$$

{\bf Acknowledgement}

 The authors  acknowledge financial support by the Ministry of Education,
Youth and Sports of the Czech Republic, project no. CZ.02.1.01/0.0/0.0/16\_019/0000778.

\end{document}